\documentclass[11pt]{article}

\usepackage[a4paper,margin=2.4cm]{geometry}
\usepackage{amsmath,amssymb,bm}
\usepackage{booktabs,array,multirow}
\usepackage{graphicx}
\usepackage{xcolor}
\usepackage{epstopdf}
\usepackage{dsfont,makecell}
\usepackage{natbib}
\usepackage{hyperref}

\title{Systemic risk in empirical interbank networks: disentangling balance sheets and network effects}

\author{Alessandro Ferracci$^1$ and Giulio Cimini$^{2,3,4,\ast}$\\
\small $^1$Department of Economics and Law, Sapienza University of Rome, 00161 Rome (Italy)\\
\small $^2$Department of Physics and INFN, University of Rome Tor Vergata, 00133 Rome (Italy)\\
\small $^3$ Enrico Fermi Research Center, 00184 Rome (Italy)\\
\small $^4$ IMT School for Advanced Studies, 55100 Lucca (Italy)\\
\small $^\ast$ giulio.cimini@roma2.infn.it}

\date{}

\begin{document}
\maketitle

\begin{abstract}
We study how the level of systemic risk empirically observed in an interbank network differs from the expected risk implied by the balance sheets composition of the participating banks, abstracting from the specific configuration of bilateral exposures.
To this end, using different contagion models (i.e., interbank asset valuation functions), we measure observed systemic risk on e-MID network data from 2005 to 2012 and compare it with the expected risk generated by a null model: an ensemble of interbank networks whose links represent possible market clearing configurations. 
This counterfactual is obtained through a maximum-entropy approach that preserves banks' balance sheet variables and diversification patterns, while randomizing the configuration of exposures.
We find that the aggregate levels of observed and expected systemic risks are generally consistent but diverge significantly during turbulent periods -- specifically following the default of Lehman Brothers (2009) and the VLTRO implementation by the ECB (2012). 
At the individual level, however, banks often appear more or less risky than implied by their balance sheet alone, due to their position in the network. 
Our results show, on the one hand, that balance sheet information embedded in a suitable maximum-entropy framework provides reliable estimates of aggregate systemic risk; on the other hand, they highlight the importance of detailed network information for accurate stress testing of individual banks, particularly after systemic events.
\end{abstract}

\newpage

\section{Introduction} \label{sec:introduction}
	
Research about contagion in financial networks has attracted substantial interest in the last twenty years, especially following the Global Financial Crisis of 2007/08 (GFC) (see for instance \cite{Bougheas2015,Battiston2016,glasserman2016,Gai2019,Bardoscia2021,Jackson2021,ELLIOTT2021105157}). 
Although systemic risk can arise in several contexts \citep{benoit2017}, this stream of literature has focused on understanding how spillovers generated by interconnectedness among financial institutions influence aggregate outcomes. A key driver for this scientific effort has been the demand by regulators and policymakers for a deeper understanding of how interconnections affect systemic risk and financial stability \citep{Haldane2009,Yellen2013}. 
As a result, metrics of financial interconnectedness have been included into the Basel Framework for the evaluation of Global Systemically Important Banks.

Within the broader literature on financial interconnections, particular attention has been devoted to interbank networks, defined by considering each loan among two banks as a directed link between the lender and the borrower, thereby translating the balance sheet structure into a network. When a shock hits one or more banks, the network structure enables distress to spread across connected institutions, potentially amplifying its effects. 
Considerable effort has thus been devoted to understanding contagion dynamics, quantifying the probability and severity of a systemic event, and overall assessing how the stability of the system depends on the structure of the network \citep{Huser2015,SIGMUND2024101223}.

Empirical research addressed interbank contagion even before the GFC. 
Several studies analyzed the possible effects of contagion on real or estimated networks in US \citep{Furfine2003}, UK \citep{Wells2004,Elsinger2005UK}, Austria \citep{Boss2004,Elsinger2006AU}, Brazil \citep{Cont2010} and Italy \citep{Mistrulli2011}. In his review, \cite{Upper2011} concluded that the risk of contagion should be taken seriously because a systemic event, even if rare, could affect a substantial part of the banking system.\footnote{Recent studies show that the risk of contagion has been steadily decreasing since the GFC \citep{battiston2015capital,Bardoscia2019}, thanks to the Basel III reforms that increased the capital requirements and introduced limits to the size of the exposures. However, studies also show that modeling only one channel of contagion severely underestimates the risks generated by interconnectedness \citep{caccioli2015overlapping,montagna2020origin}, hence the risks arising from the interbank market is still relevant overall.} 
These studies rely either on the contagion-at-default algorithm by \cite{Furfine2003}, where the value of an interbank asset is equal to its face value as long as the debtor has not defaulted, and it is equal to zero otherwise, or on the \cite{Eisenberg2001} model for clearing payments, which corresponds to a contagion-at-default with an endogenous recovery rate.
A more recent methodology to address financial contagion is the DebtRank algorithm \citep{Battiston2012, Bardoscia2015}, based on the assumption that relative shocks to equities of debtors are linearly transmitted to interbank assets of creditors. This approach has been further generalized by \cite{Barucca2020} into the NEVA (Network Valuation of Financial Assets) framework, also accounting for the presence of uncertainty on banks' external assets. 

From a more theoretical viewpoint, some authors have studied how systemic risk and stability depend on the structure of the interbank network.
Pioneers in this kind of analysis were \cite{Allen2000} and \cite{Freixas2000}.
\cite{Allen2000} considered a model in which regional banks are connected through interbank transactions and are subject to non-perfectly correlated deposit withdrawals. Even if their model is simple enough to provide a closed-form solution, they showed that contagion is highly dependent on the network structure, with the complete network perfectly spreading the shock over the entire system and the ring network being the most fragile topology. In a similar setting, \cite{Freixas2000} concluded that more links (i.e., higher network density) increase the ability of the system to absorb shocks and thus its resiliency. 
Recent advancement on the topic was given by \cite{Acemoglu2015}. Using a generalization of the Eisenberg and Noe model, they showed that the financial system displays a ``robust-yet-fragile'' structure: for small shocks, a denser network increases the resiliency, thereby confirming the results of \cite{Allen2000} and \cite{Freixas2000}; for shocks over a certain threshold instead, increasing the density of the network increases its fragility. Previous evidence of the non-monotonic effects of the connectivity was provided by \cite{Gai2010}, using contagion-at-default dynamics on Poisson random graphs (where each edge has a fixed probability to exist, independently of the other edges (\cite{ER59}). They showed that the probability of contagion changes non-monotonically with the average degree of the network: as connectivity grows, the probability of a systemic event first increases and then, after a certain threshold, gets lower until it vanishes. Similarly, \cite{BattistonJFS2012} found that individual diversification has ambiguous effects on the size of the default cascades -- depending on the distribution of banks' robustness, the size of the initial shock, and the presence of runs.  Using DebtRank dynamics, \cite{Bardoscia2017} showed that instability increases with the network density, the reason being the presence of more cycles of exposures among banks that act as risk amplifiers. 
More recently, \cite{Ramadiah2020} found that the dependence of systemic risk on the network density varies according to the type of contagion dynamics. They further considered block-structured networks, showing that risk is higher in marked core-periphery topologies due to the presence of a strongly connected set of banks that enhance the contagion process. 

A common aspect to all these works is the use of synthetic interbank networks, either derived from a model or generated through a numerical recipe. Notably, this approach allows to probe the effect of a network feature by modifying its value while leaving other characteristics unchanged -- for example, by varying the density of connections while keeping banks' balance sheets fixed. Clearly this approach cannot be applied to networks derived from empirical data, which however constitute the true test bed for financial contagion algorithms. 
In this work we propose a new framework to understand whether (and how much) the network features of an empirical interbank market influence its financial stability. The idea is to compare the systemic risk \textit{observed} for an empirical network with a benchmark that captures the risk \textit{expected} solely from the balance sheet structure of the system. We estimate the latter using a null model, the Separable Directed Enhanced Configuration Model (SDECM), a maximum-entropy construction that generates random networks while preserving, for each bank, the number of counterparties and the total volume of interbank assets and liabilities (i.e., the relevant balance sheet variables). This null allows to statistically test the hypothesis that the observed level of systemic risk is driven by the specific topological configuration of the empirical network, beyond the individual bank features encoded in balance sheets. 

We implement our framework using data on the electronic Market of Interbank Deposits (e-MID), consisting of interbank exposures between (anonymized) banks, from 2005Q1 to 2012Q3. This data allows building the empirical interbank network, as well as computing total interbank assets and liabilities of the participating banks to inform the null model. However, it contains no information about banks' equity -- which is crucial to investigate the resilience of the network. We thus derive a matching between interbank position and total equity of each bank by exploiting the relation between these variables from BankFocus data. 
We then measure systemic risk of the whole market and systemic relevance of individual institutions, using the NEVA framework \citep{Barucca2020} with the valuation functions that are frequently used in the literature, namely Furfine and DebtRank.

Our results show that the level of observed systemic risk aggregated for all banks is usually compatible with its expected value implied by the balance sheets composition of the market. However, significant deviations appear during turbulent times -- namely, the failure of Lehman Brothers (2009 network) and the introduction of the 3-year LTRO (2012 network). 
In 2009 the observed systemic risk is lower than the expected value -- a finding in line with the literature on interbank markets during the GFC \citep{Angelini2011,Afonso2011,Affinito2017}, showing that banks had increased concerns about counterparty risk, in particular after the Default of Lehman Brothers.
On the contrary, in 2012 we observe positive deviations that we argue are generated by the introduction of the 3-year LTRO, which had a profound impact on interbank markets \citep{Affinito2017,Barucca2018} and probably decreased banks' risk aversion, consistently with the risk-taking channel of monetary policy \citep{Borio2008}.
On a more granular level, instead, the knowledge of the exact position of a bank in the network is often required to assess its systemic importance (with the largest deviations of observed and expected values still observed in 2009 and 2012).

Overall, our results underline the importance of using network models and collecting data to precisely monitor contagion effects and thus ensure financial stability. Using only balance sheet information is not enough, especially when it matters the most --- that is when disruptive events are at play. Supervisors and policy makers should thus continue with their efforts in monitoring the network structure of the market and using network models for stress testing the financial system.

	\section{Methods} \label{sec:datameth}

As mentioned in the introduction, our goal is to compare the systemic risk \textit{observed} for an empirical interbank network with a benchmark, given by the risk \textit{expected} solely from the balance sheet structure of the market 
-- which we compute through the SDECM null. 
Our research question is whether the empirical link pattern amplifies or dampens risk beyond what the margins alone would imply. 
In this respect, SDECM plays a role analogous to a counterfactual in policy analysis: it provides the outcome we would expect to observe if only balance-sheet variables mattered and link patterns were random. Therefore, rejection of the null means that using only the knowledge of balance sheets (leading to SDECM simulated networks) leads to significant mis-estimation of risk, and more information should be used when performing stress tests. On the contrary, acceptance of the null means that the knowledge of balance sheets is enough to obtain risk estimates that are compatible with those observed empirically.

Our framework consists of the following steps (each detailed in the sections below):
\begin{enumerate}
\item Use e-MID data to build the empirical interbank network (one for each year), and infer banks' equities using BankFocus data.
\item Measure aggregate systemic risk of the whole market and systemic relevance of individual banks, using Furfine and DebtRank dynamics (within the NEVA formulation).
\item Use SDECM to build an ensemble of random networks, in which the number of counterparties and the total exposure of each bank are as in the empirical data.
\item Implement a statistical test by comparing \textit{observed} systemic risk (measured on empirical e-MID data) and 
\textit{expected} systemic risk (computed on the SDECM ensemble).
\end{enumerate}
	
\subsection{Data} \label{sec:data}
	
Data on interbank exposures are usually confidential and thus not easily accessible. To obtain a proxy of interbank relationships, we use data from the electronic Market of Interbank Deposits (e-MID), a trading platform of unsecured money-market loans. e-MID covers the entire domestic overnight deposit market in Italy but is also open to foreign banks.\footnote{We remark that e-MID data can be considered only a proxy for the actual direct exposures among banks. However, it is extensively used in literature as it provides valuable information on the whole structure of interbank relationships. As shown by \cite{Beaupain2011}, e-MID price data closely mirror the behavior of the Euro overnight market, with some deviations appearing during turmoil periods. Furthermore, \cite{Bargigli2015} showed that the overall interbank market, obtained by aggregating short and long maturities, closely resembles the overnight one. Based on European data that include information about several instruments, also \cite{Aldasoro2018} show that the short-term network closely resembles the total one, and that the credit claim network is an acceptable proxy for the total network. Yet, since by the mid-2010s unsecured overnight lending had nearly disappeared and was replaced by secured transactions \citep{Bardoscia2018}, our results should be interpreted as characterizing the interbank market in the 2005–2012 period.} 
The data considered in this work comprises every transaction from 2005Q1 to 2012Q3; for each contract we have information about the amount exchanged, the date and time, the interest rate, the IDs of the lender and of the borrower banks, and the contract maturity from overnight to 1 year (we remand the reader to \cite{Brandi2018} for detailed description of the data). We restrict the analysis to overnight loans that account for roughly 90\% of the total volume. \cite{Finger2013} showed that in order to have information about the underlying ``latent'' network of preferential lending relationships, data must be aggregated at least to the monthly level. For consistency with the other data source used in the paper (see below), we aggregate data at the yearly level by summing up all interbank loans between two counterparties in a given year. 

Thus, for each year, we have a network of $n$ nodes, corresponding to the banks operating in the market. Each gross loan from bank $i$ to bank $j$ is represented by a directed weighted link $w_{ij}$ connecting the two nodes $i$ and $j$. These links are grouped into a $n\times n$ matrix $W$ of interbank exposures.
From $W$ we can derive the $n\times n$ adjacency matrix $B$, whose binary element $b_{ij}$ is equal to $1$ if $w_{ij}>0$ and equal to $0$ otherwise.
These matrices encode node-specific properties: the out-degree $k_i^{out}$ and the in-degree $k_i^{in}$ represent the number of counterparties that are borrowers and lenders of bank $i$: 
	\begin{align}
		k_i^{out}=\sum_{j=1} ^n b_{ij} &&  k_i^{in}=\sum_{j=1}^n b_{ji}
	\end{align}
while the out-strength $s_i^{out}$ and the in-strength $s_i^{in}$ represent the total amount of interbank assets $A_i$ and liabilities $L_i$ of $i$:
	\begin{align}
		s_i^{out}\equiv A_i=\sum_{j=1} ^n w_{ij} &&  s_i^{in}\equiv L_i=\sum_{j=1}^n w_{ji}
	\end{align}
	
\begin{figure}[t]
		\centering
		\includegraphics[width=0.75\textwidth]{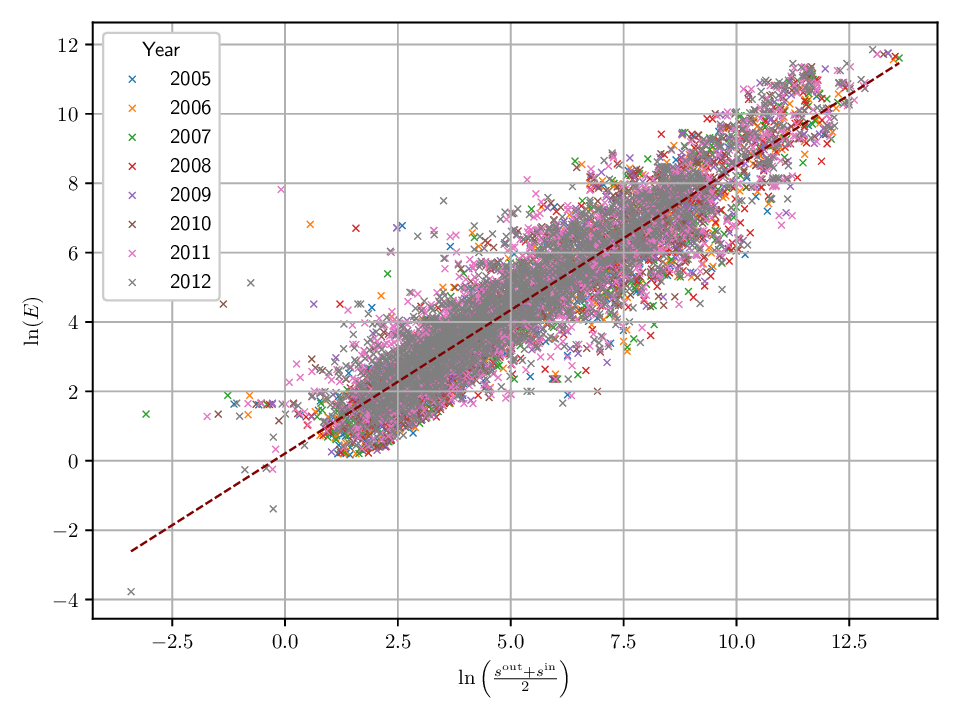}
		\caption{Scatter plot of the yearly logarithmic values of average interbank position and equity for all the banks included in the BankFocus dataset. As these quantities change through the sample, the regression is run on a yearly basis. The dashed line has slope 0.83 and represents the average value obtained by the yearly regressions.}
		\label{fig:BF}
	\end{figure}

To run the contagion dynamics on the network, we also need information about the equity of each bank. Unfortunately, e-MID data does not contain this information; moreover, banks are anonymized in the dataset, hence it is also impossible to get this information from other sources. To overcome this limitation, we employ data from BankFocus about the balance sheet compositions of financial institutions in the Euro area. BankFocus provides yearly information on equity, loans and advances to banks, and deposits from banks between 2005 and 2012 for a set of 2366 financial institutions.
Using such data we perform, for each year, a simple linear regression between the logarithm of equity and the logarithm of the average between interbank assets and liabilities. 
These two variables show a strong linear relationship, with an average Pearson correlation coefficient of 0.94 and $R^2$ of 0.88 from the regressions.
\footnote{We use the average of interbank assets and liabilities in the regression because these quantities are less symmetrical in e-MID (where some small banks act only either as lender or borrower) than in BankFocus data. 
Thus, using the average provides a stable proxy for interbank intensity that is comparable across all banks. This choice also increases the quality of the outcome: as robustness check, we have performed regressions where we include net position as an additional variable, estimate separate slopes by net-ratio quartiles, and run median regressions. Results remain qualitatively unchanged (the $R^2$ actually decreases to 0.75 due to the presence of additional variables).
Note however that our aim is not to identify a causal relationship between interbank positions and equity, but to obtain a predictive correlation that can be applied consistently to the anonymized e-MID dataset. Since e-MID contains only interbank transactions, the regression based on BankFocus data provides a practical mapping from the available variables to equity values. For this reason, we do not add additional BankFocus controls, as they cannot be matched to the anonymized e-MID nodes. In this context, the fact that fixed effects absorb the influence of omitted variables is not a drawback but an advantage, as it improves predictive accuracy given the available information.} 
The validity of this relationship is also shown in the scatter plot of Figure \ref{fig:BF}. 
	
The resulting correlation between interbank position and equity of each bank in BankFocus can then be used to infer the average equity of banks that have a given interbank exposure in e-MID. To do so, we rescale e-MID interbank positions to match the BankFocus average and use the parameters obtained from the linear regression to generate a fictional value for the yearly equity of each bank in the e-MID network (see Appendix \ref{cons_check} for more details and robustness checks).

    Note that the use of fictional equities does not hinder our analysis, since the same values are imputed in both the empirical e-MID network and the null model. In other words, the composition of balance sheets remains the same in the two scenarios, leaving the link pattern as the only changing variable. This allows us to disentangle between the inherent instability of the system due to balance sheets and the risks generated by the observed network configuration. Indeed, we are not interested in the absolute value of systemic risk but in the difference between observed and expected scenarios.

    We are now in a position to describe the balance sheet of bank $i$ in the e-MID data through the following equation
	\begin{equation}
		A_i + N_i^E= L_i + E_i 
	\end{equation}
	where $N_i^E$ represents the total amount of net external assets, computed as the residual of the accounting identity.\footnote{e-MID data record transactions at the level of individual institutions, and anonymization prevents consolidation within banking groups. Using unconsolidated exposures follows the standard practice in the literature, as it preserves bottom-up contagion mechanisms that consolidation may obscure. Difficulties in a subsidiary can affect third parties before any group-level support intervenes, and even when support is eventually provided, the need for intervention is itself informative, as such implicit guarantees were an important contagion channel during the global financial crisis. Our focus is therefore on assessing the systemic impact and vulnerability of individual institutions rather than group-level exposures.}

	\subsection{Systemic risk measures} \label{systemic} 
	
	 We then use the NEVA framework \citep{Barucca2020} to assess the level of systemic risk in the network. In the model, after a shock hits the system, each bank re-evaluates its interbank claims and updates the value of its equity accordingly. Subsequently, other rounds of assets valuation are performed until the equity values converge. The iterative map for the equity of each bank can be computed from the balance sheet identity as follows:
	\begin{equation}\label{neva}
		E_i(t+1)= N_i^E + \sum_{j=1}^N w_{ij}\mathbb{V}(E_j(t))-L_i
	\end{equation}
In this expression, $\mathbb{V}$ represents the valuation function used to re-evaluate interbank assets, which is assumed to be homogeneous among all banks and will be defined below according to different hypotheses, while $t$ stands for the valuation round (that is, the iteration step of the algorithm). 
	The NEVA framework is general enough to encompass different contagion dynamics. In order to compare with previous literature, we use a valuation function that generates contagion on default dynamics, making NEVA equivalent to the \cite{Furfine2003} algorithm. However, given that the hypothesis of contagion only upon default can be particularly strong \citep{Upper2011}, we will also use valuation functions that allow for distress contagion, corresponding to DebtRank in both its linear and nonlinear forms \citep{Bardoscia2015, Bardoscia2016}.
	
\begin{figure}[t]
		\centering
	   \includegraphics[width=0.75\textwidth]{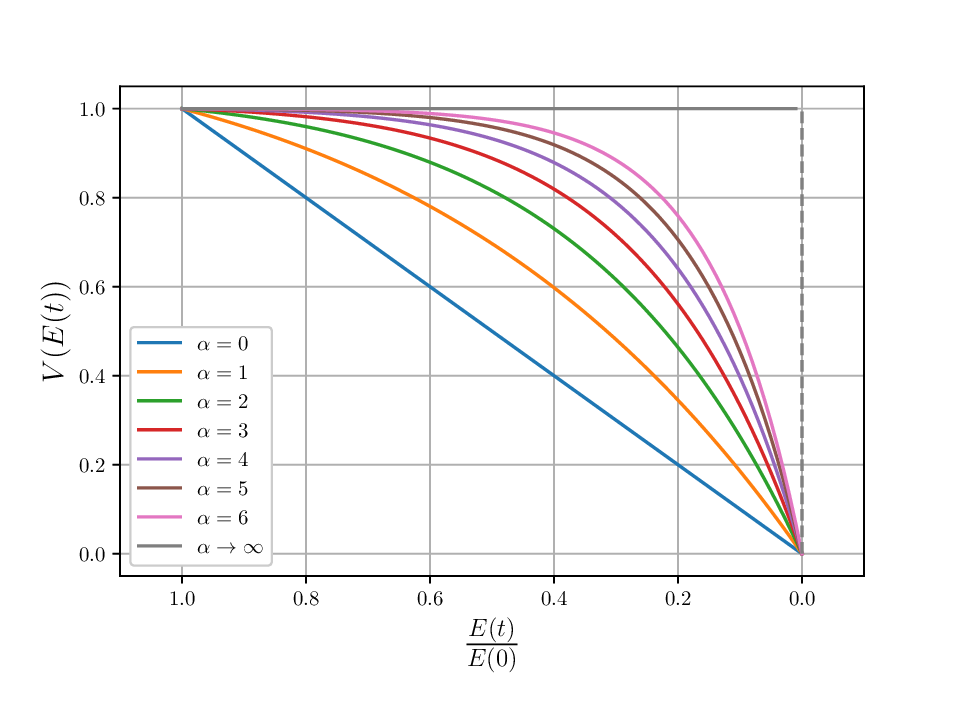}
	   \caption{Behavior of the asset valuation function of eq. \eqref{non_lin} in terms of equity loss for the counterparty, for different values of the nonlinearity parameter $\alpha$ ranging from $\alpha=0$ (linear DebtRank) to $\alpha\to\infty$ (Furfine, or contagion on default).}
		\label{fig:nonlin}
	\end{figure}

	\begin{itemize}
	\item The Furfine algorithm works essentially in three steps. First, a bank $i$ (or a set of banks) fails by assumption. Then contagion propagates through the network, and any other bank $j$ that lent to bank $i$ suffers a loss $(1-R)w_{ji}$, where $0\le R< 1$ is the recovery rate (i.e., the proportion that the lending bank is able to recover after the default of the counterparty). If $(1-R)w_{ji}>E_j$, bank $j$ defaults and a new round of propagation starts. Such dynamics can be obtained in NEVA through the following valuation function: 
	\begin{equation}\label{furfine}
		\mathbb{V}_{F}(E_j(t))=
		\mathds{1} _{E_j(t)\geq 0}+R\mathds{1} _{E_j(t)< 0}
	\end{equation}
	where $\mathds{1}$ is an indicator function taking the value of $1$ if the relative condition is met.
	
	\item Even when no default has occurred, the market value of a loan can drop if the borrower bank is in distress. To model this situation, we use a valuation function compatible with the linear DebtRank dynamics
	\begin{equation}\label{lin_dr}
		\mathbb{V}_{DR}(E_j(t))=
		min \left[ max \left( \frac{E_j(t)}{E_j(0)},0 \right) ,1 \right]
	\end{equation}
	In this case, the equity losses experienced by a bank are always passed to its creditors: 
	each percentage drop of equity is reflected in an equal reduction of interbank claims value.
	
	\item Finally, we also use a valuation function that maps to the nonlinear DebtRank model
	\begin{equation}\label{non_lin}
		\mathbb{V}_{DR(\alpha)}(E_j(t))=
		\left( \mathbb{V}_{[DR]}(E_j(t))-1 \right) e^{- \alpha \, \mathbb{V}_{[DR]}(E_j(t))}
		+ 1 
	\end{equation}
	In this case, distress is propagated nonlinearly according to the parameter $\alpha$, which sets the functional dependence of the default probability on the relative equity loss: 
    the higher this value, the larger the shock banks can withstand before their probability of default increases. The extreme case $\alpha=0$ corresponds to linear DebtRank and $\alpha\to\infty$ to the Furfine algorithm (see Figure \ref{fig:nonlin}).
\end{itemize}

    We remark that while Furfine and DebtRank can be considered as different forms of the valuation function within the same (NEVA) framework, the economic mechanisms they represent are rather different. The option structure of debt contracts, such as interbank loans, implies that losses take a zero-one form: either the bank defaults and the value of its liabilities drops to zero (with the caveat of what can be recovered at the end of the bankruptcy procedure), or it does not default, and the liabilities are honored by the banks. According to this scheme, contagion-at-default can be a more suitable approach to analyze short-term contracts, such as overnight ones.
  
    DebtRank instead looks at the market value of interbank claims as a function of banks' equity losses, and is thus more closely related to credit valuation for long-term loans that accounts for the reduction in value generated by a higher probability of default. Indeed, while liabilities maintain their face value, an interbank asset is worth less if the debtor bank is hit by a shock or faces higher risk, hence the creditor bank books \textit{potential} losses and its equity decreases, transmitting such losses further. Thereby, DebtRank can be interpreted as a measure of the value of interbank exposures that considers the risk embedded in the network structure of credit claims.
	
	\subsection{Null Model} \label{null}
	
Maximum entropy network models derived from Statistical Physics and Information Theory enable generating an ensemble of networks that are maximally random but satisfy (on average) some constraints, which thus represent the sufficient statistics of the model (in the sense that they explain any property of the generated networks) \citep{jaynes1957information}.
Therefore, they are widely used to run statistical tests, i.e., to assess the significance of network patterns under specific null hypotheses. 
For example, if property $Y$ measured on an empirical network is also reproduced by a maximum entropy model constraining property $X$ of the original network, then we immediately know that $Y$ is not an independent feature but just a statistical consequence of $X$ \citep{squartini2011analytical,Cimini2019}. 
Applications of the methodology close to our scope are provided by \cite{Squartini2013} and \cite{Bargigli2015}, who tested whether the interbank market is characterized by an anomalous presence of network motifs that can be interpreted as early warning signs of systemic crises. Using similar methods, \cite{Hatzopoulos2015} tested the e-MID market network for the presence of preferential lending relationships. Other examples are provided by \cite{Gualdi2016} about overlapping portfolios, \cite{Garlaschelli2004} about world trade and \cite{Mancini2026} for production networks. 
 
In this paper we employ a maximum entropy model that allows generating random networks while keeping fixed the total amount of interbank lending and borrowing \textit{and} the level of diversification of each bank. The latter variable (node-level connectivity) is included as constraint since it deeply affects the outcome of contagion, as extensively shown by the literature discussed in the Introduction.\footnote{Entropy maximization without any degree or connectivity constraints would spread exposures as evenly as possible, resulting in a very dense network where banks are more diversified than they actually are (\cite{Squartini2018}).} 
Comparing empirical data with this maximum entropy construction thus allows us to assess whether systemic risk is driven solely by the balance sheet composition of the system, or to quantify the additional contribution of the specific link pattern observed in the empirical network -- while controlling for variables known to have strong impact on systemic risk.
 
In practice, the model we use is the \textit{Separable Directed Enhanced Configuration model} (SDECM), which we formulate building on its undirected counterpart introduced by \citet{Gabrielli2019}.
    The model uses as constraints the number of counterparties (borrowers $k_i^{out}$ and lenders $k_i^{in}$) and the total exposure (volumes lent $s_i^{out}=A_i$ and borrowed $s_i^{in}=L_i$) for each bank $i$. In order to achieve a good compromise between mathematical rigor and computational efficiency, the model is based on a two-step entropy maximization procedure: firstly, degree constraints are used to obtain the probability of the adjacency matrix configurations; then, strength constraints are used to obtain the probability density of the link weights conditional to a given binary configuration. Finally, likelihood maximization allows us to adjust model parameters so that values of model degrees and strengths match empirical observations \citep{squartini2011analytical}. For the full mathematical details of the model, see Appendix \ref{SDECM}.
    	
	Using SDECM and constraint values taken from the empirical e-MID data, we generate for each year an ensemble of 1000 networks that shall constitute our null model. With respect to the empirical network, each element in the ensemble has the same number $n$ of banks; on average over this ensemble, each bank has the same number of counterparties and the same total interbank borrowing and lending; apart from satisfying these constraints, transactions between banks and their volumes are placed randomly, such that each network in the ensemble represents a different market clearing configuration. This, plus the use of fixed equity values simulated from BankFocus, ensures that the level of interconnectedness and the balance sheets composition of the null model coincides (on average) with that of the empirical network.
	
	We use the contagion models previously described on each generated network to obtain average values and standard deviations for various systemic risk indicators. We shall refer to these quantities as the \textit{expected} systemic risk metrics conditional to the balance sheets structure of the market. The comparison of such expected indicators with the ones \textit{observed} in the real e-MID network will allow us to disentangle the risk inherent to the balance sheets with the risk due to the specific configuration of the network.
	Furthermore, the SDECM enables us to remain as agnostic as possible with respect to the network formation process of the interbank market. Given that the matching mechanisms of the market are likely changing during the time span of the data, with the SDECM we assume the simplest formation process, i.e., the one driven by pure randomness. This choice gives us the possibility to observe if banks' choices of counterparties change throughout our sample and if those different choices entail different levels of systemic risk.

	\section{Results}
	
	We now consider two different settings concerning the initial shock. Firstly in Section \ref{sec:aggr} we apply a systemic shock to all banks and analyze the aggregate level of systemic risk. Then in Section \ref{sec:rel} we assess the systemic relevance of individual banks by letting them default, one by one.

	\subsection{Aggregate systemic risk} \label{sec:aggr}
	
	Here we study the effect of a small exogenous initial shock distributed on the entire network. This shock is simulated, for each bank $i$, as a reduction of the net external assets $N_i^E$ in a way proportional to the equity $E_i(0)$. That is:
	\begin{equation}
		E_i(1)= (1-\lambda) E_i(0)
	\end{equation}
	where the parameter $\lambda$ controls the shock size and is homogeneous among all banks. We then assess the losses due to the reverberation of this shock using eq. \eqref{neva}. We quantify the effects of contagion as the mean relative loss of equity in the system:
    \begin{equation}
    H= \frac{\sum_{i=1}^n \left[E_i(1)-E_i(t^*)\right]}{\sum_{j=1}^n E_j(0)}
	\end{equation}
	where $E_i(t^*)$ represents the equity value at the end of the shock propagation dynamics. In this way, we measure the impact of the diffusion and feedback effects generated through the re-evaluation of interbank assets without considering the direct effect of the initial shock.
    $H$ varies between $0$, when shocks do not propagate, and $1-\lambda$, when all banks have defaulted.
	
	\begin{figure}[t]
		\centering
		\includegraphics[width=\textwidth]{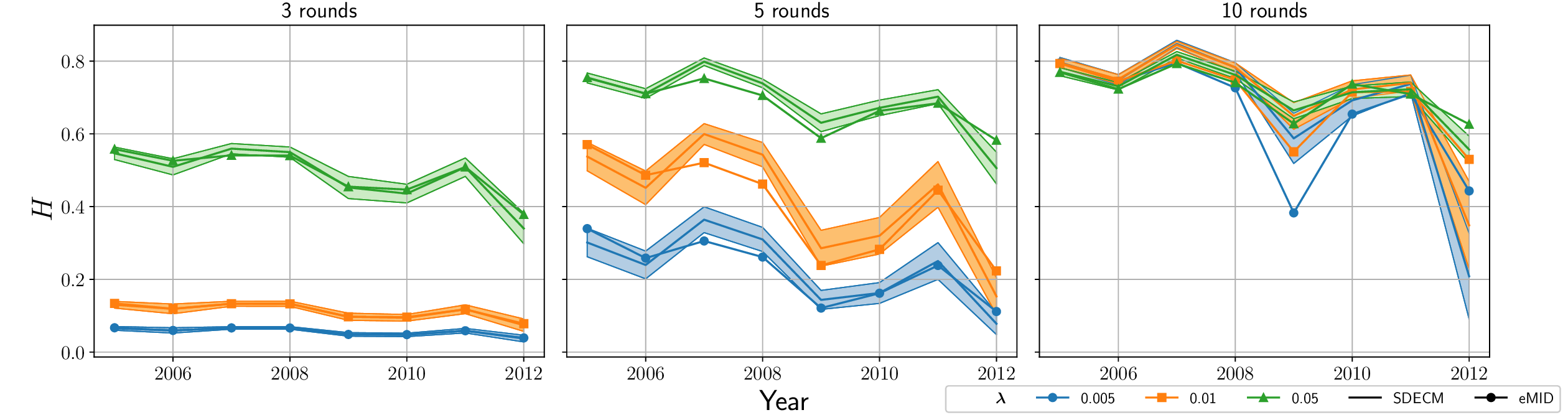}
		\caption{Linear DebtRank dynamics of eq. \eqref{lin_dr}: aggregate systemic risk $H$ in the various yearly networks, for some sample iteration steps of the dynamics (3, 5 and 10 rounds in the left, middle and right panel respectively) and varying initial shock (0.5\% in blue, 1\% in orange and 5\% in green). At 10 rounds of propagation, the system has almost reached convergence. Values observed in the empirical networks are indicated with dots connected with solid lines, whereas SDECM expected values are represented with solid lines and shaded areas (indicating ensemble mean and standard deviation).}
		\label{fig:lin_dr}
	\end{figure}
	
	In this section we only use DebtRank valuation functions, due to the lack of initial defaults that could trigger Furfine dynamics. 
	For the linear DebtRank of eq. \eqref{lin_dr}, Figure \ref{fig:lin_dr} shows expected and observed aggregate systemic risk for some sample iteration steps of the algorithm, allowing us to evaluate the speed at which contagion spreads through the system. After 10 rounds the system has almost reached convergence.
    We firstly notice how the aggregate risk $H$ can be different in the various yearly networks considered. The contagion dynamic runs faster for initial shocks of larger magnitude, but the stationary values of $H$ are usually very close between each other.
	Concerning the relation between observed and expected systemic risk, we observe a good agreement during the early stages of the dynamics (the observed value of $H$ falls within the confidence region of one standard deviation $\sigma$ for the expected $H$). Slight deviations start to appear for intermediate iteration steps (particularly for 2007 and 2008), while at convergence we observe only two  economically significant deviations. In the 2009 network, observed systemic risk is 14.8\% (3.3$\sigma$) and 12.7\% (3$\sigma$) lower than expected for $\lambda$ equal to 0.005 and 0.01, even if the deviation reduces only to 3\% (0.8$\sigma$) for the higher value 0.05. The 2012 network has the opposite behavior, with an observed $H$ that is 50\%, 25\%, and 13\% (1.18, 1.27, and 1.94 $\sigma$) higher than expected, respectively for $\lambda$ equal to 0.005, 0.01 and 0.05.

\begin{figure}[t]
		\centering
		\includegraphics[width=\textwidth]{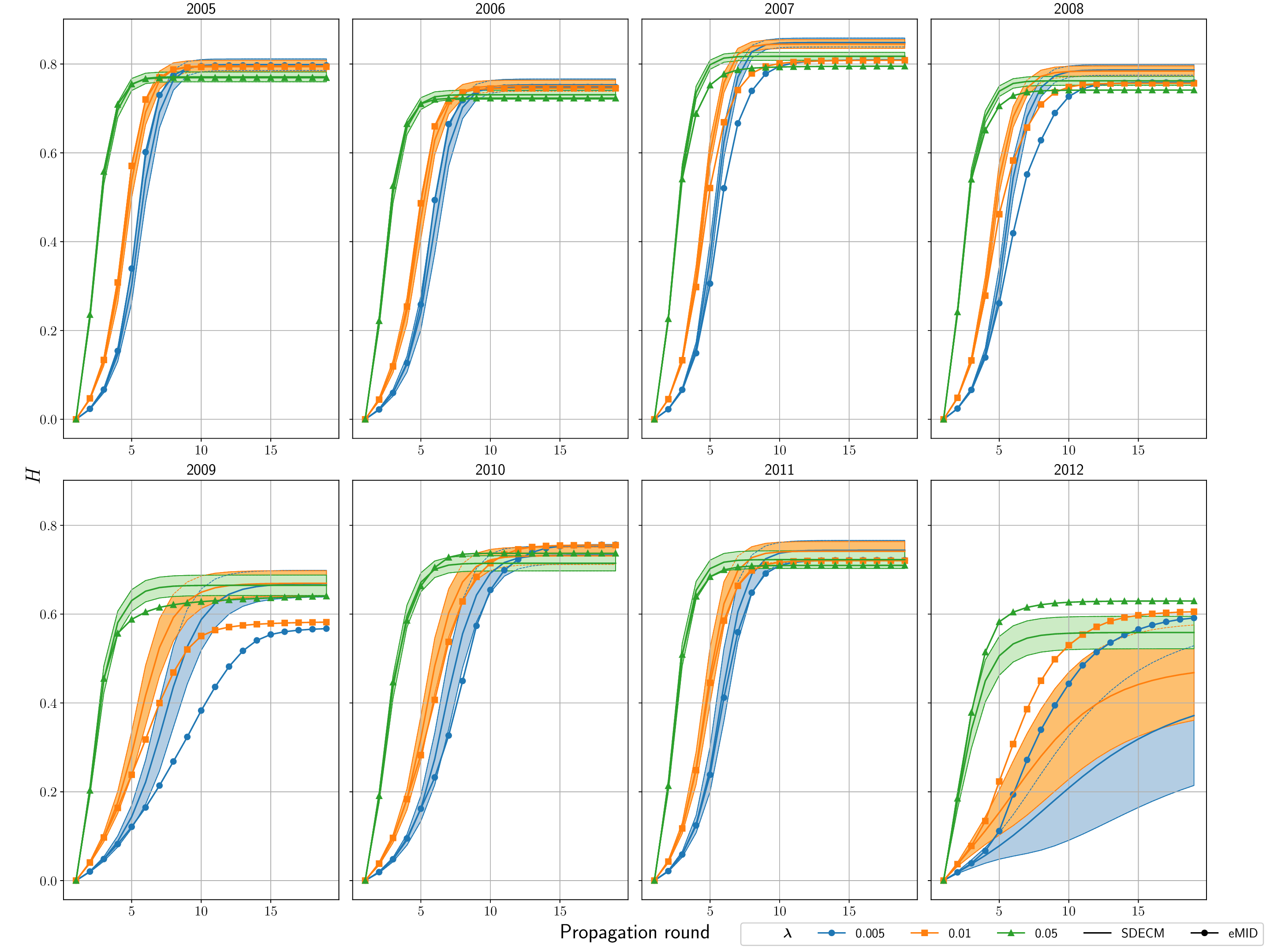}
        \caption{Linear DebtRank dynamics of eq. \eqref{lin_dr}: evolution of aggregate systemic risk $H$ as a function of the iteration step, for the various yearly networks considered and varying initial shock $\lambda$. Observed and expected values are indicated respectively with solid dots (connected by lines) and line plus shaded area (indicating ensemble mean and standard deviation).}
		\label{fig:steps}
	\end{figure}

	A clearer picture of the contagion propagation can be obtained from Figure \ref{fig:steps} where, for each yearly network, we plot how the value of $H$ grows with the iteration step (i.e., the round of contagion). Table \ref{table:linear} in Appendix \ref{sec:tables_aggr} provides numerical details about the Figure.
	The only yearly networks for which observed and expected systemic risk are statistically different for all the values of $\lambda$ are 2007, 2008, 2009 and 2012.
	We confirm that in 2007 and 2008 the observed systemic risk is slightly lower than expected, with an economically significant difference only during the first iterations of the algorithm. However, 2009 stands out for the magnitude of the deviation, especially for smaller initial shocks: observed systemic risk is on average 15.3\% (2.54$\sigma$) lower than what is expected from the balance sheets distribution. In 2012 we observe an opposite behavior with the e-MID network generating, on average, a relative equity loss that is 47.2\% higher (1.28$\sigma$) than expected. All the negative deviations tend to disappear for large initial shocks ($\lambda=0.05$) but remain significant for the 2012 network, with observed systemic risk that is on average 12\% (1.73$\sigma$) higher than the expected value.
    
\begin{figure}[t]
		\centering
		\includegraphics[width=\textwidth]{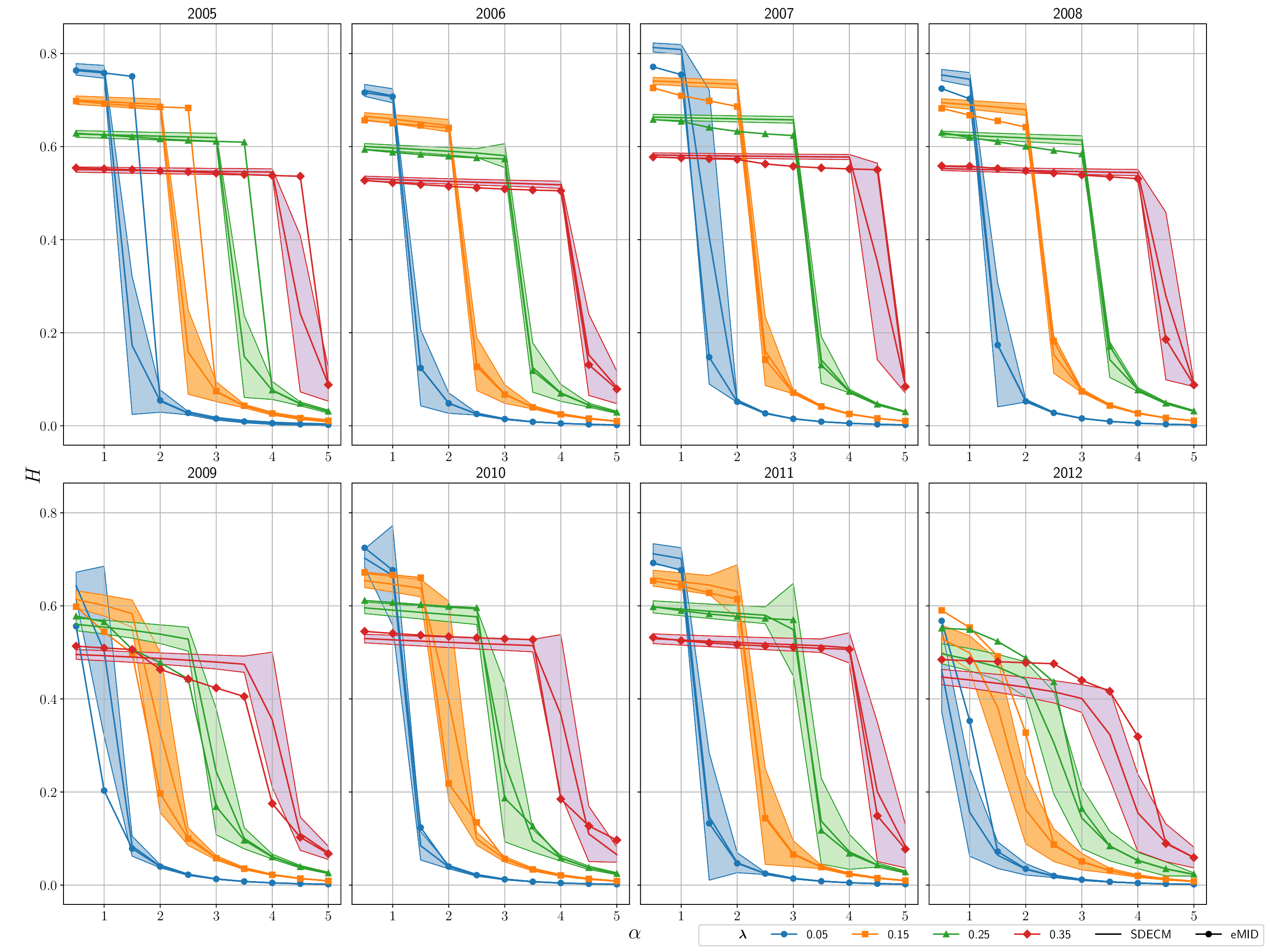}
		\caption{Nonlinear DebtRank dynamics of eq. \eqref{non_lin}: stationary level of aggregate systemic risk $H$ as a function of the nonlinearity parameter $\alpha$, for the various yearly networks considered and varying initial shock $\lambda$. Observed and expected values are indicated respectively with solid dots (connected by lines) and line plus shaded area (indicating ensemble mean and standard deviation). We recall that $\alpha=0$ corresponds to linear DebtRank, while $\alpha\to\infty$ leads in principle to contagion on default.}
		\label{fig:nonlin_aggr}
	\end{figure}

	For the case of nonlinear DebtRank of eq. \eqref{non_lin}, Figure \ref{fig:nonlin_aggr} shows the stationary values of the average relative equity loss as a function of the claims re-evaluation factor $\alpha$, while Table \ref{table:nonlinear} contains the related numerical details. 
    The bulk of deviations between observed and expected risk is consistent with the previous analysis. The difference is statistically significant and negative for 2007, 2008 and especially 2009, for values of the initial shock $\lambda\le 0.25$ (small positive deviations are observed for $\alpha<2$ and large shock $\lambda=0.35$ in 2009\footnote{Some deviations are also present in 2006 and 2010, yet they do not have a general pattern but are specific to certain combinations of $\lambda$ and $\alpha$ values. These results are likely due to the large instability of $H$ for intermediate values of the nonlinearity parameter $\alpha$, as shown also by \cite{Ramadiah2020}.}). 
    Large and positive significant differences are instead observed in  2012 for all values of non-linearity and initial shock parameters.

	\subsection{Systemic relevance} \label{sec:rel}
	
	We now analyze the systemic relevance of each bank $i$ by using two indicators: impact $I_i$ and vulnerability $V_i$ \citep{Cimini2016}.
	The impact $I_i$ measures the relative amount of equity lost by the system as a consequence of the initial default of bank $i$
	\begin{equation}
		I_i= 1 - \frac{\sum_{j\neq i}^n E_j(t^*| E_i(1)=0)}{\sum_{j\neq i}^n E_j (0)}
	\end{equation}
	while the vulnerability $V_i$ quantifies the average equity loss of bank $i$ due to the default of another bank in the network
	\begin{equation}
		V_i = \frac{1}{n-1} \sum_{j\neq i}^n \left[1-\frac{E_i(t^*| E_j(1)=0)}{E_i(0)}\right]
	\end{equation}
	In both cases, the condition $t^*| E_i(1)=0$ indicates the final step of a dynamics with initial condition given by the default of bank $i$, and no shocks to all other banks.
	
	\begin{figure}[t]
		\centering
		\includegraphics[width=\textwidth]{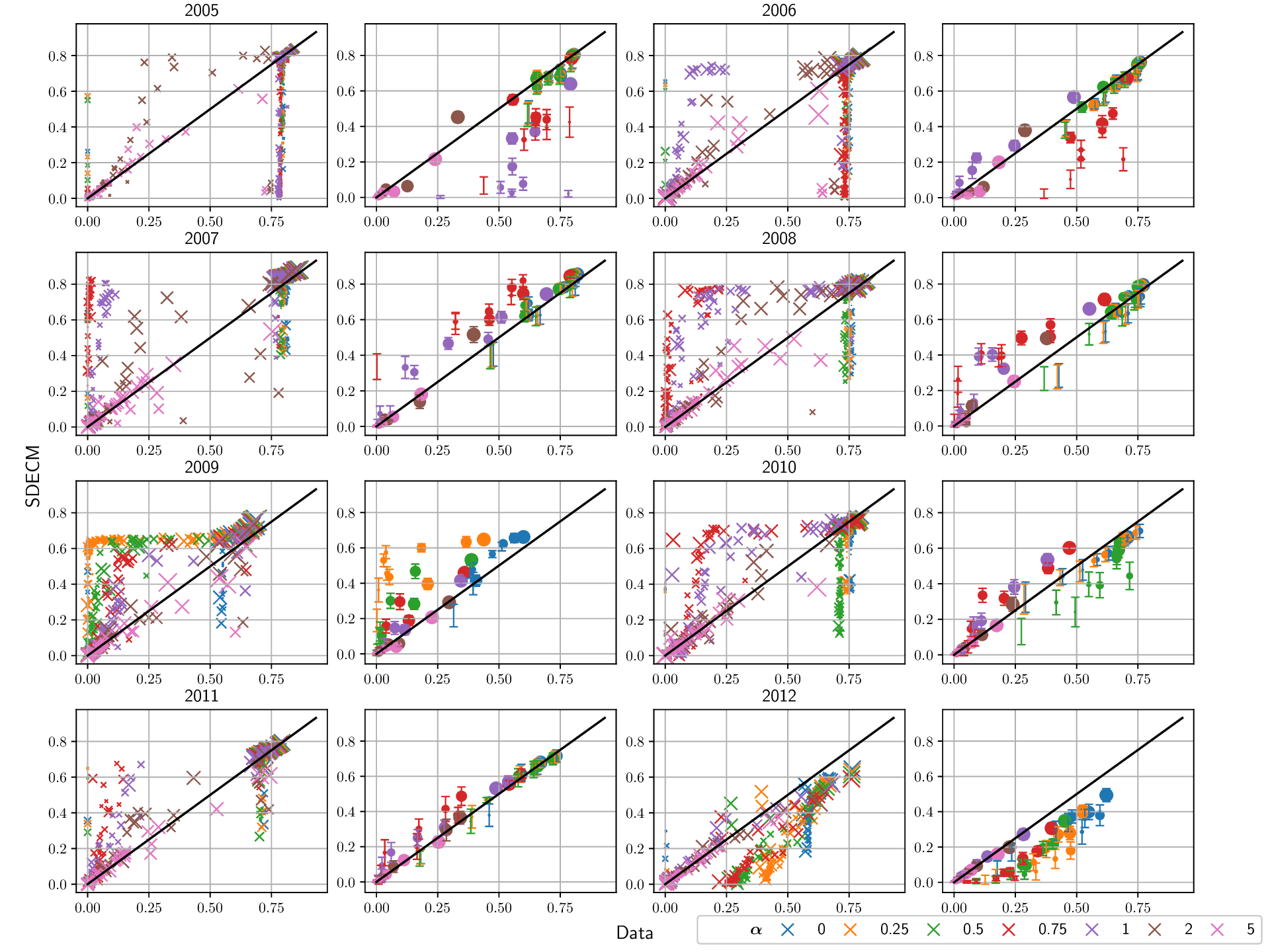}
		\caption{Scatter plot of observed (Data) and expected (SDECM) impact of individual banks in the system for the various yearly networks considered. Different values of the nonlinearity parameter $\alpha$ are displayed with varying colors. For each year, the plot on the left shows each bank with a different point, while the plot on the right shows the average value of impact in the equity decile group.}
		\label{fig:impact}
	\end{figure}
	
\begin{figure}[t]
		\centering
		\includegraphics[width=\textwidth]{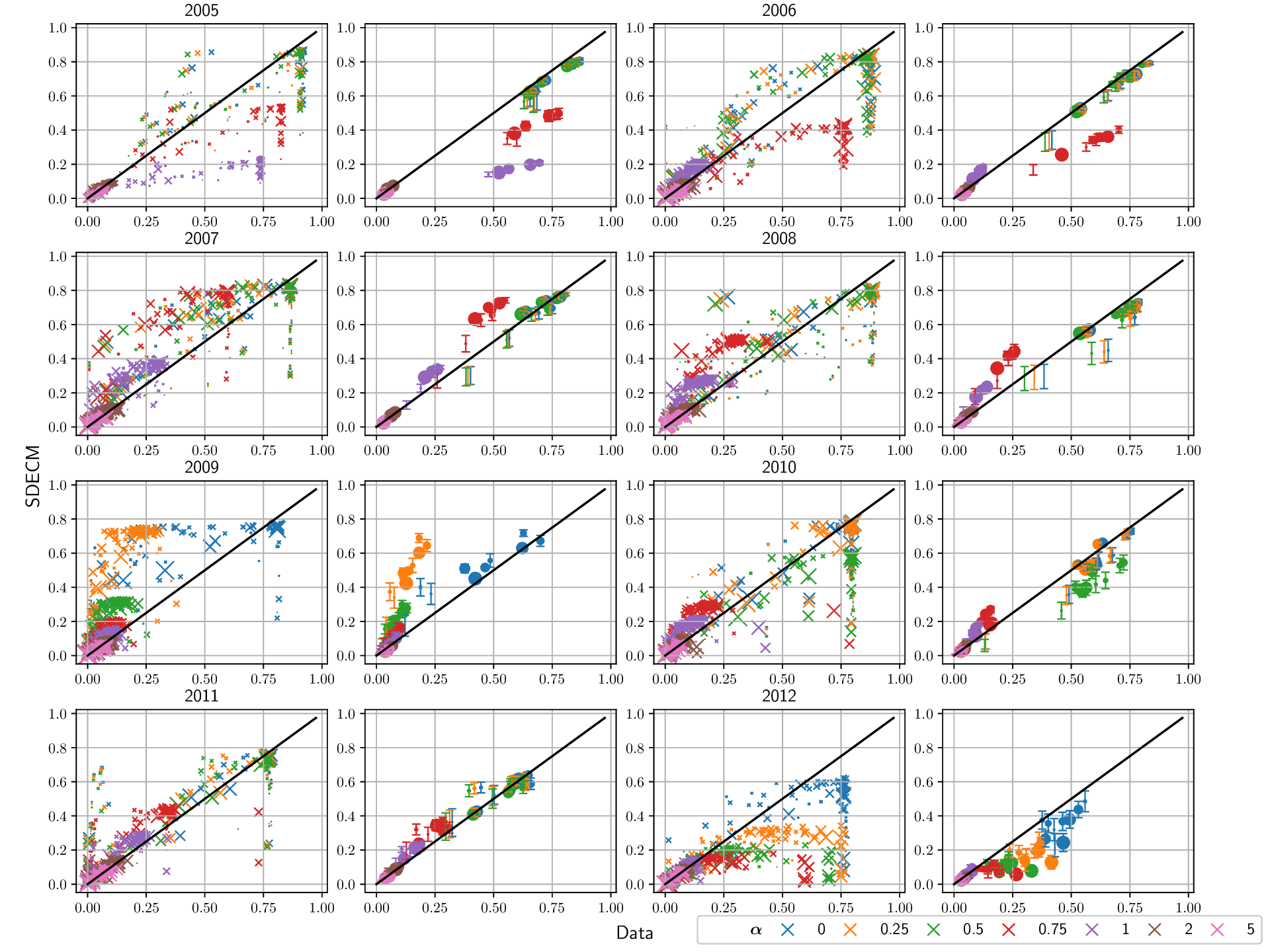}
		\caption{Scatter plot of observed (Data) and expected (SDECM) vulnerability of individual banks in the system, for the various yearly networks considered. Different values of the nonlinearity parameter $\alpha$ are displayed with varying colors. For each year, the plot on the left shows each bank with a different point, while the plot on the right shows the average value of vulnerability in the equity decile group.}
		\label{fig:vuln}
	\end{figure}

\begin{figure}[t]
        \includegraphics[width=\textwidth]{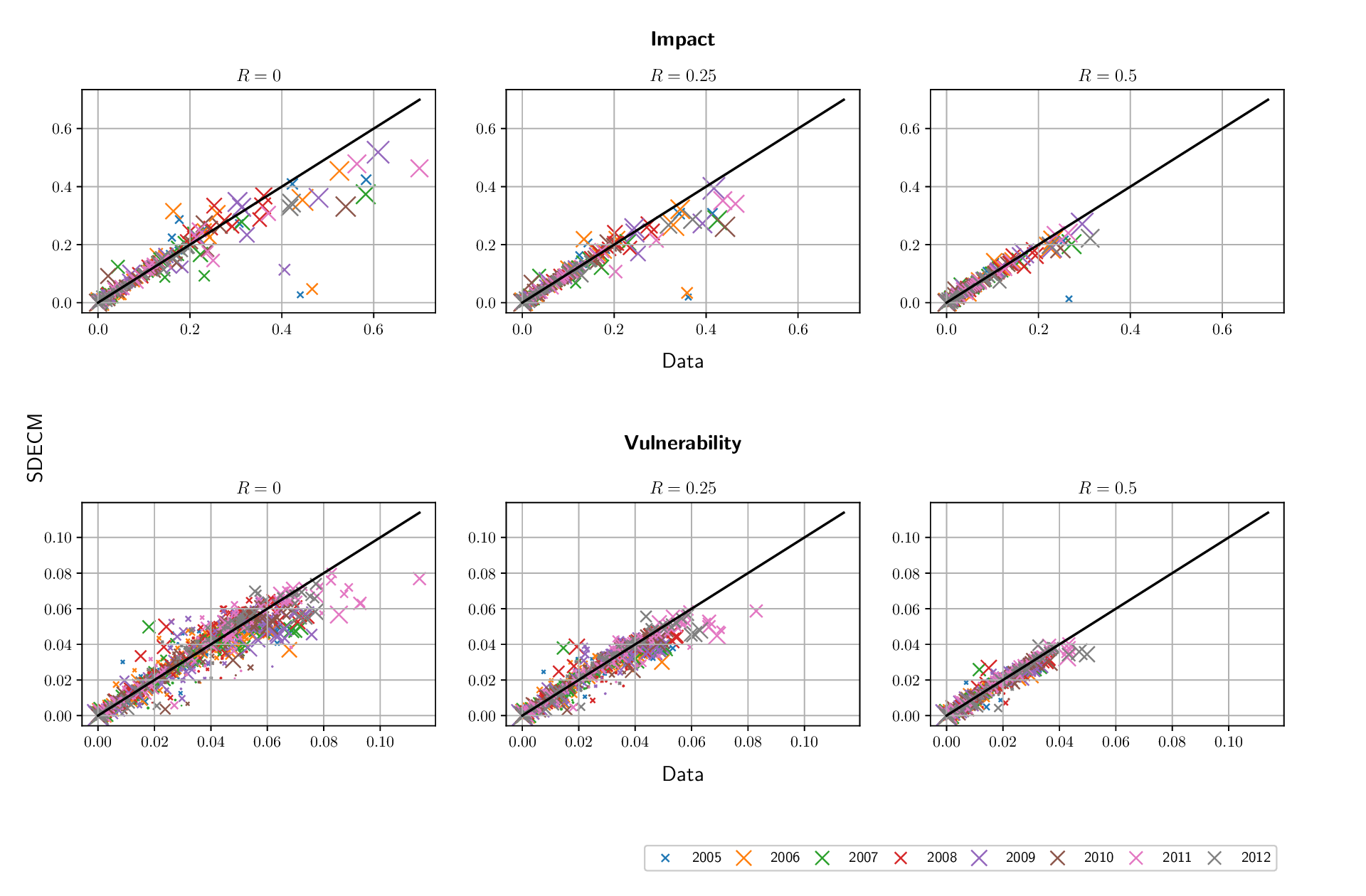}
		\caption{Scatter plot of observed (Data) and expected (SDECM) impact (top) and vulnerability (bottom) of individual banks in the system, for the Furfine dynamics of eq. \eqref{furfine} with different recovery rates $R$. Different colors denote values related to each yearly network.}
		\label{fig:furf}
	\end{figure}

Figures \ref{fig:impact} and \ref{fig:vuln} show respectively the impact and vulnerability of each bank in the various yearly networks. 
Each pair of panels refers to a specific year, the first showing the scatter plot of these quantities as computed on the data (horizontal axis) and in the SDECM ensemble (vertical axis), with colors denoting the different values of the nonlinearity parameter $\alpha$ of the DebtRank dynamics. The companion panel is structured in the same way, but values refer to averages computed on group of banks defined by splitting banks' equity distribution into deciles. The black solid line marks the identity: if empirical and model values were similar, they should lie on this bisector. Instead we observe sensible deviations (both negative and positive), especially when the nonlinearity of the valuation function is low -- that is, when banks' probabilities of default increase as soon as their equity starts decreasing.
This suggests that the position of a bank in the network matters, especially when even small losses impact the default probability and thus the value of exposures.
Figure \ref{fig:furf} confirms this observation: when the Furfine valuation function is used, deviations are nearly non-existent in all years.
	
Second, averaging banks in equity classes reduces the magnitude of the deviations, confirming the result of the previous section that the aggregate level of systemic risk can be inferred from balance sheet structures. Again, this last observation is valid only for the years when there are no extreme market disruptions. Even after averaging banks in equity classes, in 2009 deviations from the null model are clustered above the diagonal line where expected and observed systemic relevance are equal, indicating that banks are less risky than what their balance sheet would imply. In 2012 deviations are instead positive.
	
	\subsection{Discussion}	 
	We argue that the most significant deviations we observed are due to two disruptive events for the interbank market, namely the default of Lehman Brothers in September 2008 and the introduction of the 3-year Longer-Term Refinancing Operations (LTRO) by the ECB. 
	
	This observation is supported by the results of \cite{Affinito2017}. Using supervisory data from the Bank of Italy, they showed that those two events changed the network structure of the Italian interbank market. They found that the collapse of Lehman Brothers reduced the connectivity of the interbank network while, in some specifications of the model, the liquidity crisis of August 2007 is statistically insignificant. On the contrary, the announcement of the 3-year LTRO by the ECB had the opposite effect. The operations shifted the distribution of the centralities towards the most connected banks and thus to ``smoother liquidity circulation among banks''.
	With another methodology, \cite{Barucca2018} confirmed the impact of the 3-year LTRO on e-MID. They analyzed the network structure of the market from 2010 to 2014 and showed that the 3-year LTRO had a profound impact on the banks' trading behavior and thus on the network configuration. Historical lenders exited the market, and borrowing banks and new institutions stepped in taking their place, sensibly changing the network structure.
	
	The different signs of the deviations caused by the two events can be explained by the different consequences that they had on banks' behavior. \cite{Angelini2011} showed that after the liquidity crisis of August 2007, interest rates started to reflect the creditworthiness of the borrowing counterparty. Unfortunately, their analysis stops at December 2008 and thus does not help explain the results presented in this paper. However, \cite{Afonso2011}, through an analysis of the overnight Federal Funds market, found that the default of Lehman Brothers increased concerns in the market for counterparty risk, reducing liquidity and increasing costs for weaker banks. The larger impact of Lehman's default is consistent with \cite{Affinito2017} and with the deviations from the expected systemic risk that we detect in our analysis. The negative deviations that we observe in the 2009 panels of Figures \ref{fig:impact} and \ref{fig:vuln} suggest that banks have been able to reduce not only their direct counterparty risk but also to internalize part of the externalities that are generated in a network setting. Another non-trivial result is that this behavior emerges at the aggregate level, as shown in Figures \ref{fig:steps} and \ref{fig:nonlin_aggr}, with lower observed systemic risk. 
	
	On the contrary, the introduction of the 3-year LTROs could have impacted the interbank market consistently with the risk-taking channel of monetary policy \citep{Borio2008}. Our results suggest that the central bank's provision of long-term liquidity has affected the interbank market, possibly by increasing banks' willingness to take on risk and by reducing the perceived risk of contagion. The reaction of the banking system has thus generated an increased systemic risk, both at the individual (2012 panels of Figures \ref{fig:impact}-\ref{fig:vuln}) and at the aggregate level (Figures \ref{fig:steps}-\ref{fig:nonlin_aggr}).
	
	Notice how these results are consistent with the network formation mechanism that the null model implies. In the SDECM, the probability of a connection between two banks is proportional to the number of counterparties and total interbank assets and liabilities. Therefore the model describes a matching process in which individual institutions choose their counterparties independently from their probability of default. 
	The evidence from \cite{Angelini2011} and \cite{Afonso2011} suggests that counterparty risk became more relevant after the disruptions in the interbank market, making the SDECM unable to capture the systemic risk observed in the network. On the contrary, the introduction of the 3-year LTRO could have heterogeneously reduced risk aversion, thereby introducing additional features in the matching mechanism of the market and generating more systemic risk than what would be expected given the balance sheet composition of the system.
    It must be noted that the model used in this paper, while being able to detect statistically significant deviations, is not able to test for causality and further research is thus needed to uncover the effect of those events in the network formation process.	

	Moreover, our results highlight the importance of considering the exact network structure when the model used for systemic risk evaluation entails distress contagion dynamics: when the nonlinearity of DebtRank ($\alpha$) is high and with the Furfine valuation function, the deviations disappear. This suggests that, when the model approaches contagion at default, the exact network structure is less relevant. These results corroborate the analysis of \cite{glasserman2015likely}, who find that network feedbacks with the Eisenberg-Noe model increase contagion only by 1.75\%. On the contrary, distress contagion enhances the importance of the feedback effects on the network structure, as shown by the large deviations in some yearly networks.\footnote{Unfortunately \cite{glasserman2015likely} do not provide numerical details on their distress contagion extension of the Eisenberg-Noe framework.}

	\section{Related literature}	
	
	Closely related to our study are the ones of \cite{Krause2019} and \cite{Diem2020}, both based on DebtRank. Using Monte Carlo network generation, \cite{Krause2019} showed that it is possible to create financial networks with minimal systemic risk while maintaining fixed balance sheet structures. Another significant result of their analysis is that it is possible to derive approximate information about the systemic risk of a financial institution by using specific node properties, notably the interbank leverage. \cite{Diem2020} also performed a systemic risk optimization through mixed-integer linear programming: They fixed interbank assets and liabilities and optimized the network structure to obtain minimal systemic risk. 
	They showed that sensible risk reductions are possible, especially for small and medium-sized banks, by concentrating lending and funding over a minimal set of connections. 
	Our study differs from both these works since, in our exercise, we also fix the degrees of each institution: banks' diversification is kept constant, and thus concentration limits are respected. We also have a different goal: we do not aim to find the network structure of minimal systemic risk; rather, we aim to quantify how much systemic risk is due to the observed pattern of links, while controlling for the size of the interbank exposures and the level of interconnectedness of each bank -- which are known to have large effects on the risk of contagion.
    
	Another closely related study is \cite{glasserman2015likely}, who derive explicit bounds for the magnitude of network contagion while using only data about asset size, leverage and total interbank liabilities, together with minimal information about the connectivity of the network (that is, without exact knowledge about the underlying network structure). There are several significant differences between their approach and ours. First, their model does not use information about the number of counterparties of each bank, while our measures take it into account. Second, their analysis is model-dependent, as it relies on the Eisenberg-Noe framework; on the contrary, our approach can be applied to any existing model. Third, their measure allows to compute the maximum systemic risk generated by an interbank network rather than the expected one. Fourth and most importantly, they measure the contribution of the network structure by comparing this bound with a network-less system. On the contrary, in our analysis, we study the difference between the observed networks and a null model, given by the average of an ensemble of fictional networks. Our results regarding the contribution of the specific network structure are comparable to their analysis in the case of a contagion-at-default algorithm. Conversely, when the model entails distress contagion, relevant differences arise. 

A particularly relevant antecedent is \cite{Mistrulli2011}, who analyzes contagion-at-default in the Italian interbank market using observed supervisory bilateral exposures. As benchmark, he employs a classical iterative proportional fitting algorithm to derive a maximum-entropy (MaxEnt) interbank matrix that preserves total interbank assets and liabilities of each bank. However, this construction implies a complete, over-diversified network where all pairs of banks are connected. 
Comparing contagion simulations on observed versus benchmark networks, he finds that the MaxEnt approach typically underestimates contagion, but under certain conditions (depending on network structure, recovery rates, and capitalization) it may also overestimate it. 
We instead construct our benchmark networks with the Separable Directed Enhanced Configuration Model (SDECM), which also preserves each bank’s diversification. 
This improvement allows us to disentangle balance-sheet and connectivity effects from the specific configuration of bilateral exposures, so to statistically assess whether the observed systemic risk departs from balance-sheet expectations. 
Our analysis is also broader in scope, as we verify that our conclusions are robust across alternative contagion mechanisms and over time, which allows us to pinpoint when deviations become material during systemic episodes.

Lastly, we briefly discuss the use of maximum entropy models in financial network studies (we remand to  \cite{Cimini2019} and \cite{Bardoscia2021} for more details). Because they are maximally uncommittal with respect to the information imposed as constraints, they have been successfully used not only for statistical hypothesis testing the role of the constraints in shaping the empirical network (the route we follow in this paper), but also as unbiased \emph{reconstruction methods} to recover missing bilateral exposures from limited information and thus overcome the confidential nature of financial transactions (see \cite{Squartini2018} for a recent review on the topic). 
Notably, the horse-race by \cite{ANAND2018107} showed that the heuristic maximum entropy construction (\emph{density-corrected fitness model}) proposed by \cite{Cimini2015} outperforms competing recipes (e.g., \cite{Mistrulli2011,drehmann2013measuring,halaj2015modelling,anand2015filling}) in reconstructing interbank networks. 
SDECM \citep{Gabrielli2019} improves on \cite{Cimini2015} in two aspects: by explicitly constraining degrees (assumed as known, rather than derived from a fitness ansatz), and by exploiting the full distribution of link weights, rather than their average values only.
	
	\bigskip
	
	\section{Conclusion} \label{sec:conclusion}

	In this paper we contribute to the growing literature about the effects of the network structure on systemic risk, by introducing the notion of expected systemic risk, i.e. the level of systemic risk that is due to the market's balance sheets but not to a specific network configuration. This quantity is obtained using an appropriate maximum entropy network model constraining relevant balance sheet variables. 
	We compute observed and expected systemic risk, using the Network Asset Valuation framework \citep{Barucca2020}, in a dataset of interbank exposures from the e-MID market, augmented with equity information from BankFocus data.
	
	The results show that aggregate systemic risk, measured through linear and nonlinear DebtRank, is usually well explained by the configuration of balance sheets. However, significant deviations are present in the 2009 and 2012 networks.
	Conversely, measures of systemic importance at the individual bank level display substantial differences, 
    with clusters of banks that show both positive and negative deviations for DebtRank, while the results from the Furfine algorithm are more consistent with the null model. This suggests that some banks are riskier than what is implied by their balance sheet because of their position in the network. 
	The lower is the shock absorption capacity of banks (i.e., the faster the probability of default increases with equity losses), the stronger is the effect of the network positioning.
	In the limiting case of contagion on default, nearly all the deviations disappear. Moreover, averaging banks' systemic relevance in equity deciles decreases the magnitude of the deviations, except for 2009 and 2012.
	
	The negative deviations of aggregate and individual systemic risk in 2009 could signal that, after the default of Lehman Brothers in September 2008, banks collectively started to minimize their counterparty risk in response to the increased concerns about possible bankruptcies. While this could be an obvious result at the individual level, the emergence of an aggregate systemic risk reduction in a network context is not trivial. The result points towards the existence of an endogenous mechanism that allows banks to reduce their risk of being affected by contagion while also reducing systemic risk for the entire system.
	The same reasoning can be applied to the 2012 network, with the 3-year LTRO increasing banks' willingness to take on risk, reducing the perceived counterparty risk, and thus increasing the scope of contagion. While the model does not allow us to draw a causal relationship between the observed deviations and the two crisis events, we note that these observations are in line with the existing literature on interbank markets.
	
	Furthermore, our results corroborate the existing literature on interbank contagion: the network structure matters, particularly during systemic events.
	These observations demonstrate that it is essential to include networks in stress testing models and collect data to measure the scope of contagion accurately. 
	However, further research is needed to prove an exact causal relation among those events and systemic risk and to uncover the network formation processes that generate the observed effects. 
Particularly useful could be understanding which network features are responsible for the divergence of observed and expected systemic risk.\footnote{We have already tested two different empirical strategies: i) measuring how observed risk varies as a function of balance sheets and network
properties, and ii) measuring how the deviation between observed and expected risk depends on the above variables.
Unfortunately we have not been able to detect clear patterns, especially because null risk values absorb most of the explanatory power, hence we decided to leave this analysis out of the paper for future work.} 
For instance, one could study how risk depends on banks’ diversification, which can be varied in a controlled way using proper SDECM specifications. Overall, a deeper understanding of those mechanisms would be valuable for designing incentive policies that are able to reduce the instability of the network with minimal effects on banks' interbank lending and funding choices. 

    We remark that our analysis refers to the period 2005–2012, when unsecured overnight exposures were a substantial component of interbank activity and e-MID was representative of the Italian interbank market \citep{Bargigli2015}. After the Global Financial Crisis and the European sovereign debt crisis, structural changes occurred: by the mid-2010s unsecured overnight lending had nearly disappeared and was replaced by secured transactions \citep{Bardoscia2018}. Our results should therefore be interpreted as characterizing the financial system as it was in the 2005–2012 period. At the same time, the statistical validation framework we propose could, with appropriate modifications, be applied to more recent markets, including those dominated by secured lending.

\section*{Acknowledgements}
The authors are grateful to Marco Bardoscia, Stefano Battiston, Gabriele Tedeschi, the participants of the DISEI-Workshop "Heterogeneity, Evolution and Networks in Economics" and two anonymous referees for their useful comments.
This work is supported by the following projects: ‘RENet - Reconstructing economic networks: from physics to machine learning and back’, MUR PRIN 2022MTBB22 funded by European Union – Next Generation EU; ‘C2T - From Crises to Theory: towards a science of resilience and recovery for economic and financial systems’, MUR PRIN PNRR P2022E93B8 funded by European Union – Next Generation EU.

\section*{Competing Interests}
The authors declare no relevant financial or non-financial competing interests to report.

\bibliographystyle{apalike}


\appendix

\section{Robustness checks on equity imputation} \label{cons_check}

\begin{figure}[!h]
		\centering
            \includegraphics[width=\textwidth]{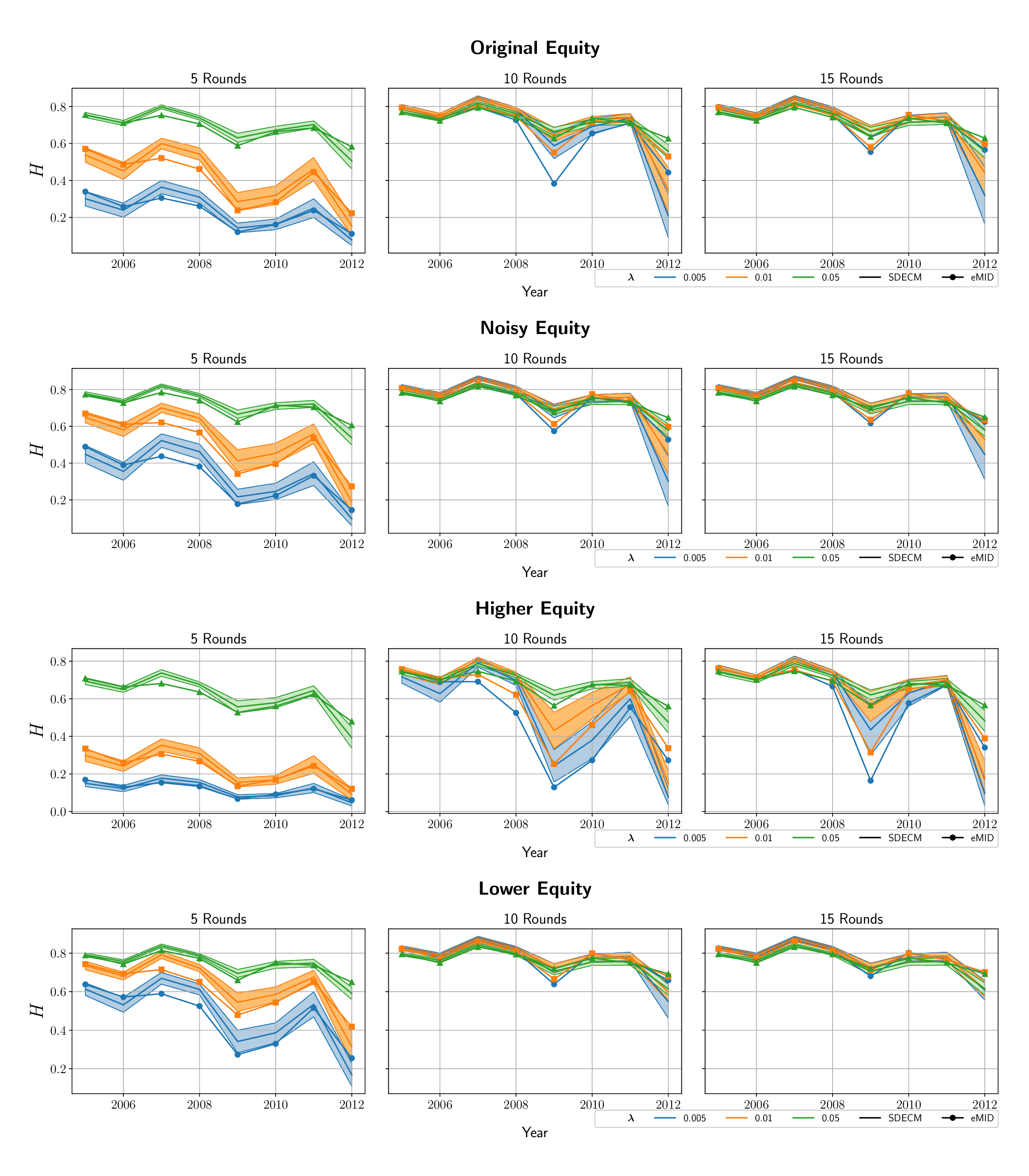}
		\caption{Linear DebtRank dynamics of eq. \eqref{lin_dr}: aggregate systemic risk $H$ in the various yearly networks, for sample iteration steps of the dynamics and varying initial shock $\lambda$ (same of Figure \ref{fig:lin_dr} in the main text). Values observed in the empirical networks are indicated with dots connected with solid lines, whereas SDECM expected values are represented with solid lines and shaded areas (indicating ensemble mean and standard deviation). The four rows of panels refer to the different equity imputation scenarios.}
		\label{fig:cons_check}
	\end{figure}

As described in the main text, we impute equity values to banks from their interbank positions through the regression:
\begin{equation}
\ln E_i = \alpha + \beta \ln \left(\frac{s_i^{out}+s_i^{in}}{2}\right) + \gamma_i
\end{equation}
Here we perform robustness checks by adopting three different scenarios for the distribution of equity among banks:
\begin{itemize}
\item Noisy equity: we add white noise on the regression coefficients before computing equities
\item Higher equity: the estimated equity of all banks is increased by 25\%
\item Lower equity: the estimated equity of all banks is decreased by 25\%
\end{itemize}
In all scenarios, predicted equities are used in the contagion dynamics for both the empirical and counterfactual networks. 
Figure \ref{fig:cons_check} shows how the results of Figure \ref{fig:lin_dr} change with these scenarios. Adding random noise does not change the picture significantly. Linearly scaling equity values does lead to different $H$ values (higher/lower risk for lower/higher equities, respectively), but our claim that null values are compatible with empirical ones, except during crises, remains consistent.

\clearpage
\pagebreak

\section{Separable Directed Enhanced Configuration Model} \label{SDECM}

Here we describe the formal steps that define the Separable Directed Enhanced Configuration Model (SDECM). We remand the reader to \cite{Gabrielli2019} for an extended presentation of the model in the simplest case of undirected networks, and to \cite{Cimini2019} for a general introduction to maximum entropy models of networks.

We start by considering a directed weighted network of $n$ nodes, denoting by ${\cal V}$ the set of all ordered node pairs so that $|{\cal V}|=n(n-1)$. The network is univocally defined through a configuration ${\cal C}=(B,W)$ composed by: the binary adjacency matrix $B=\{b_{ij}\}_{(i,j)\in{\cal V}}$ that defines the set of existing links $(i,j)\in {\cal L}\subseteq{\cal V}$ (i.e., the set of node pairs with $b_{ij}=1$), and the weighted adjacency matrix $W=\{w_{ij}\}_{(i,j)\in{\cal V}}$ that defines the set of weights $\{w_{ij}\}_{(i,j)\in {\cal L}}$ associated to them. Our goal is to construct an ensemble $\Omega$ of such networks that are maximally random, except for 4$n$ properties that are constrained on average: 
the out-degree $k_i^{out}=\sum_{j\in\cal V} b_{ij}$ and in-degree $k_i^{in}=\sum_{j\in\cal V} b_{ji}$ plus the out-strength $s_i^{out}=\sum_{j\in\cal L} w_{ij}$ and in-strength $s_i^{in}=\sum_{j\in\cal L} w_{ji}$, for each node $i$. 
The probability distribution $P({\cal C})$, that is, the occurrence probability of configuration ${\cal C}$ within the ensemble $\Omega$, is obtained by maximizing the Shannon entropy $S=-\sum_{{\cal C}\in\Omega}P(\cal C)\log P(\cal C)$ subject to the normalization condition plus the above-mentioned set of constraints. 

In principle the degrees and strengths constraints can be imposed simultaneously; however we use a two-step entropy maximization procedure that leads to an easier numerical implementation:
\begin{enumerate}
	\item We first constrain the average in/out-degrees only, obtaining the probability of the binary configuration $B$:
	\begin{equation}
		\pi(B)=\prod_{(i,j)\in{\cal V}}\frac{e^{-(\alpha^{out}_i+\alpha^{in}_j) b_{ij}}}{1+e^{-(\alpha^{out}_i+\alpha^{in}_j)}}
		\label{P-A-TS}
	\end{equation}
	where $\alpha_i^{out}$ and $\alpha_i^{in}$ are the Lagrange multipliers associated to the out/in-degree of node $i$.
	\item Then, for each $B$, we constrain the average in/out-strengths, obtaining the probability density of the link weights conditional to $B$
	\begin{equation}
		q(W_{{\cal L}})=\prod_{(i,j)\in{\cal L}}(\beta_i^{out}+\beta_j^{in})e^{-(\beta_i^{out}+\beta_j^{in})w_{ij}}
	\end{equation}
	where $\beta_i^{out}$ and $\beta_i^{in}$ are the Lagrange multipliers associated to the out/in-strength of $i$.
\end{enumerate}
Overall the SDECM is defined by the joint probability distribution
$P(B,W)=\pi(B)q(W_{{\cal L}})$. Finally to obtain a model for a specific empirical network configuration ${\cal C}^*$ (in our case, a yearly e-MID network), we maximize the likelihood of ${\cal C}^*$ in the ensemble. This corresponds to finding the values of the Lagrange multipliers that satisfy, $\forall i$:
\begin{equation}
	\begin{cases}
		[k_i^{out}]^*=\sum_{j:(i,j)\in{\cal V}}[1+e^{(\alpha^{out}_i+\alpha^{in}_j)}]^{-1}\\
		[k_i^{in}]^*=\sum_{j:(i,j)\in{\cal V}}[1+e^{(\alpha^{in}_i+\alpha^{out}_j)}]^{-1}\\
		[s_i^{out}]^*=\sum_{j:(i,j)\in{\cal L}}(\beta_i^{out}+\beta_j^{in})^{-1}[1+e^{(\alpha^{out}_i+\alpha^{in}_j)}]^{-1}\\
		[s_i^{in}]^*=\sum_{j:(i,j)\in{\cal L}}(\beta_i^{in}+\beta_j^{out})^{-1}[1+e^{(\alpha^{in}_i+\alpha^{out}_j)}]^{-1}
	\end{cases}\label{constr-SECM}
\end{equation}
where $[k_i^{out}]^*$, $[k_i^{in}]^*$, $[s_i^{out}]^*$, $[s_i^{in}]^*$ are the numerical values that constraints assume on the empirical network ${\cal C}^*$. \footnote{Note that using a non-uniform prior for the parameters would lead to a  Bayesian-like framework \citep{Gandy2017}.}

\pagebreak

\section{Results tables} \label{sec:tables_aggr}

\begin{table}[h!]
	\centering
	\resizebox{0.8\textwidth}{!}{
		\begin{tabular}{llcccccccc}
			\toprule
			& \textbf{Year} &                                           2005 &                                           2006 &                                                2007 &                                                2008 &                                                2009 &                                             2010 &                                           2011 &                                              2012 \\
			$\bm{\lambda}$ & $\bm{t}$ &                                                &                                                &                                                     &                                                     &                                                     &                                                  &                                                &                                                   \\
			\midrule
			\multirow{5}{*}{\textbf{0.005}} & \textbf{3 } &    \thead{$0.026$ \\ \textit{0.328} \\ (69.3)} &  \thead{$-0.001$ \\ \textit{-0.005} \\ (61.1)} &         \thead{$0.001$ \\ \textit{0.023} \\ (53.3)} &       \thead{$-0.002$ \\ \textit{-0.043} \\ (51.4)} &       \thead{$-0.002$ \\ \textit{-0.019} \\ (57.2)} &      \thead{$0.017$ \\ \textit{0.165} \\ (64.0)} &  \thead{$-0.004$ \\ \textit{-0.039} \\ (53.1)} &         \thead{$0.049$ \\ \textit{0.2} \\ (72.4)} \\
			& \textbf{5 } &    \thead{$0.126$ \\ \textit{0.971} \\ (85.1)} &    \thead{$0.079$ \\ \textit{0.496} \\ (76.6)} &   \thead{$-0.161^{**}$ \\ \textit{-1.649} \\ (3.8)} &    \thead{$-0.156^{*}$ \\ \textit{-1.459} \\ (5.1)} &       \thead{$-0.157$ \\ \textit{-0.861} \\ (17.7)} &    \thead{$-0.004$ \\ \textit{-0.024} \\ (55.0)} &  \thead{$-0.049$ \\ \textit{-0.242} \\ (44.3)} &        \thead{$0.43$ \\ \textit{1.124} \\ (89.8)} \\
			& \textbf{10} &  \thead{$-0.001$ \\ \textit{-0.085} \\ (44.4)} &  \thead{$-0.005$ \\ \textit{-0.262} \\ (37.9)} &    \thead{$-0.061^{***}$ \\ \textit{-5.1} \\ (0.0)} &  \thead{$-0.071^{***}$ \\ \textit{-4.324} \\ (0.0)} &  \thead{$-0.349^{***}$ \\ \textit{-2.967} \\ (0.7)} &    \thead{$-0.055$ \\ \textit{-0.877} \\ (16.3)} &   \thead{$-0.038$ \\ \textit{-1.15} \\ (11.3)} &   \thead{$1.126^{*}$ \\ \textit{1.991} \\ (93.7)} \\
			& \textbf{20} &  \thead{$-0.002$ \\ \textit{-0.131} \\ (42.6)} &    \thead{$-0.007$ \\ \textit{-0.4} \\ (32.4)} &  \thead{$-0.046^{***}$ \\ \textit{-3.894} \\ (0.0)} &   \thead{$-0.036^{**}$ \\ \textit{-2.379} \\ (1.7)} &  \thead{$-0.151^{***}$ \\ \textit{-3.379} \\ (0.1)} &      \thead{$0.032$ \\ \textit{1.173} \\ (87.9)} &   \thead{$-0.03$ \\ \textit{-0.999} \\ (17.3)} &   \thead{$0.558^{*}$ \\ \textit{1.357} \\ (92.0)} \\
			& \textbf{Convergence} &  \thead{$-0.002$ \\ \textit{-0.131} \\ (42.6)} &    \thead{$-0.007$ \\ \textit{-0.4} \\ (32.4)} &  \thead{$-0.045^{***}$ \\ \textit{-3.879} \\ (0.0)} &   \thead{$-0.036^{**}$ \\ \textit{-2.379} \\ (1.7)} &   \thead{$-0.148^{***}$ \\ \textit{-3.34} \\ (0.1)} &      \thead{$0.032$ \\ \textit{1.177} \\ (88.1)} &   \thead{$-0.03$ \\ \textit{-0.999} \\ (17.3)} &   \thead{$0.403^{*}$ \\ \textit{1.176} \\ (92.3)} \\
			\cline{1-10}
			\multirow{5}{*}{\textbf{0.010}} & \textbf{3 } &    \thead{$0.028$ \\ \textit{0.388} \\ (69.3)} &    \thead{$0.003$ \\ \textit{0.028} \\ (61.1)} &         \thead{$0.001$ \\ \textit{0.023} \\ (53.3)} &        \thead{$-0.002$ \\ \textit{-0.04} \\ (51.4)} &       \thead{$-0.002$ \\ \textit{-0.017} \\ (57.2)} &      \thead{$0.019$ \\ \textit{0.199} \\ (64.1)} &  \thead{$-0.003$ \\ \textit{-0.033} \\ (53.1)} &        \thead{$0.053$ \\ \textit{0.23} \\ (72.4)} \\
			& \textbf{5 } &     \thead{$0.061$ \\ \textit{0.84} \\ (80.4)} &    \thead{$0.077$ \\ \textit{0.752} \\ (78.8)} &  \thead{$-0.131^{***}$ \\ \textit{-2.754} \\ (0.6)} &   \thead{$-0.149^{**}$ \\ \textit{-2.427} \\ (1.2)} &       \thead{$-0.165$ \\ \textit{-0.956} \\ (15.4)} &    \thead{$-0.117$ \\ \textit{-0.744} \\ (23.2)} &  \thead{$-0.034$ \\ \textit{-0.253} \\ (38.9)} &    \thead{$0.456^{*}$ \\ \textit{1.38} \\ (91.6)} \\
			& \textbf{10} &  \thead{$-0.002$ \\ \textit{-0.127} \\ (42.9)} &  \thead{$-0.007$ \\ \textit{-0.403} \\ (32.3)} &  \thead{$-0.051^{***}$ \\ \textit{-4.384} \\ (0.0)} &  \thead{$-0.044^{***}$ \\ \textit{-2.919} \\ (0.4)} &  \thead{$-0.152^{***}$ \\ \textit{-2.696} \\ (0.7)} &    \thead{$-0.011$ \\ \textit{-0.321} \\ (34.2)} &  \thead{$-0.032$ \\ \textit{-1.057} \\ (14.6)} &        \thead{$0.52$ \\ \textit{1.504} \\ (89.7)} \\
			& \textbf{20} &   \thead{$-0.002$ \\ \textit{-0.13} \\ (42.8)} &   \thead{$-0.007$ \\ \textit{-0.44} \\ (31.0)} &  \thead{$-0.043^{***}$ \\ \textit{-3.707} \\ (0.0)} &   \thead{$-0.035^{**}$ \\ \textit{-2.368} \\ (1.7)} &   \thead{$-0.13^{***}$ \\ \textit{-3.063} \\ (0.1)} &      \thead{$0.033$ \\ \textit{1.231} \\ (89.8)} &  \thead{$-0.029$ \\ \textit{-0.967} \\ (18.2)} &   \thead{$0.284^{*}$ \\ \textit{1.271} \\ (93.0)} \\
			& \textbf{Convergence} &   \thead{$-0.002$ \\ \textit{-0.13} \\ (42.8)} &   \thead{$-0.007$ \\ \textit{-0.44} \\ (31.0)} &  \thead{$-0.043^{***}$ \\ \textit{-3.696} \\ (0.0)} &   \thead{$-0.035^{**}$ \\ \textit{-2.368} \\ (1.7)} &  \thead{$-0.127^{***}$ \\ \textit{-2.997} \\ (0.2)} &      \thead{$0.033$ \\ \textit{1.232} \\ (89.8)} &  \thead{$-0.029$ \\ \textit{-0.967} \\ (18.2)} &    \thead{$0.25^{*}$ \\ \textit{1.264} \\ (93.1)} \\
			\cline{1-10}
			\multirow{5}{*}{\textbf{0.050}} & \textbf{3 } &    \thead{$0.021$ \\ \textit{0.672} \\ (74.8)} &     \thead{$0.033$ \\ \textit{0.75} \\ (78.6)} &    \thead{$-0.033^{*}$ \\ \textit{-1.307} \\ (8.7)} &        \thead{$-0.017$ \\ \textit{-0.66} \\ (24.5)} &         \thead{$0.005$ \\ \textit{0.077} \\ (56.6)} &      \thead{$0.025$ \\ \textit{0.427} \\ (68.1)} &    \thead{$0.001$ \\ \textit{0.018} \\ (49.0)} &       \thead{$0.113$ \\ \textit{0.919} \\ (82.9)} \\
			& \textbf{5 } &    \thead{$0.001$ \\ \textit{0.081} \\ (51.5)} &   \thead{$-0.001$ \\ \textit{-0.03} \\ (45.2)} &  \thead{$-0.057^{***}$ \\ \textit{-4.434} \\ (0.0)} &  \thead{$-0.045^{***}$ \\ \textit{-2.787} \\ (0.4)} &   \thead{$-0.067^{**}$ \\ \textit{-1.732} \\ (3.5)} &    \thead{$-0.012$ \\ \textit{-0.385} \\ (34.7)} &  \thead{$-0.026$ \\ \textit{30-0.942} \\ (17.6)} &  \thead{$0.152^{**}$ \\ \textit{1.753} \\ (95.3)} \\
			& \textbf{10} &  \thead{$-0.002$ \\ \textit{-0.163} \\ (41.5)} &  \thead{$-0.011$ \\ \textit{-0.762} \\ (20.5)} &  \thead{$-0.029^{***}$ \\ \textit{-2.676} \\ (0.5)} &   \thead{$-0.028^{**}$ \\ \textit{-2.094} \\ (2.6)} &    \thead{$-0.055^{*}$ \\ \textit{-1.585} \\ (6.0)} &  \thead{$0.032^{*}$ \\ \textit{1.311} \\ (90.4)} &  \thead{$-0.018$ \\ \textit{-0.666} \\ (28.0)} &   \thead{$0.125^{**}$ \\ \textit{1.89} \\ (96.4)} \\
			& \textbf{20} &  \thead{$-0.002$ \\ \textit{-0.163} \\ (41.5)} &  \thead{$-0.011$ \\ \textit{-0.763} \\ (20.5)} &  \thead{$-0.027^{***}$ \\ \textit{-2.476} \\ (0.8)} &   \thead{$-0.027^{**}$ \\ \textit{-2.073} \\ (2.6)} &       \thead{$-0.036$ \\ \textit{-1.028} \\ (15.4)} &  \thead{$0.032^{*}$ \\ \textit{1.311} \\ (90.4)} &  \thead{$-0.018$ \\ \textit{-0.654} \\ (28.0)} &  \thead{$0.127^{**}$ \\ \textit{1.936} \\ (96.6)} \\
			& \textbf{Convergence} &  \thead{$-0.002$ \\ \textit{-0.163} \\ (41.5)} &  \thead{$-0.011$ \\ \textit{-0.763} \\ (20.5)} &  \thead{$-0.027^{***}$ \\ \textit{-2.473} \\ (0.8)} &   \thead{$-0.027^{**}$ \\ \textit{-2.073} \\ (2.6)} &       \thead{$-0.029$ \\ \textit{-0.844} \\ (19.6)} &  \thead{$0.032^{*}$ \\ \textit{1.311} \\ (90.4)} &  \thead{$-0.018$ \\ \textit{-0.654} \\ (28.0)} &  \thead{$0.127^{**}$ \\ \textit{1.936} \\ (96.6)} \\
			\bottomrule
	\end{tabular}}
	\caption{\scriptsize \textbf{Linear DebtRank}.
		This table reports data about the relative equity loss of the system $H$ with the linear DebtRank valuation function (Figure \ref{fig:steps}). Each row corresponds to a combination of the initial shock $\lambda$ and contagion round $t$ while the columns correspond to the yearly networks. The top element from each cell corresponds to the relative deviation of $H$ for e-MID data from its SDECM average; the middle elements in \textit{italics} correspond to the difference between e-MID and SDECM average divided by SDECM standard deviation; the bottom element in parenthesis reports the percentile position of e-MID with respect to the distribution of $H$ generated by SDECM. *, ** and *** denote the significance of the deviation and highlight if e-MID's $H$ is either in the top or bottom 10\%, 5\% or 1\% of the distribution generated by SDECM.}
	\label{table:linear}
\end{table}

\pagebreak

		\begin{table}[h!]
			\centering
			\resizebox{0.7\textwidth}{!}{
					\begin{tabular}{llcccccccccc}
				\toprule
				& \textbf{Year} &                                              2005 &                                               2006 &                                                2007 &                                                2008 &                                                2009 &                                               2010 &                                           2011 &                                               2012 \\
				$\bm{\lambda}$& $\bm{\alpha}$ &                                                   &                                                    &                                                     &                                                     &                                                     &                                                    &                                                &                                                    \\
				\midrule
				\multirow{5}[30]{*}{\textbf{0.05}} & \textbf{0.5} &     \thead{$-0.003$ \\ \textit{-0.189} \\ (40.1)} &      \thead{$-0.007$ \\ \textit{-0.374} \\ (33.2)} &  \thead{$-0.051^{***}$ \\ \textit{-4.264} \\ (0.0)} &   \thead{$-0.039^{**}$ \\ \textit{-2.465} \\ (1.4)} &  \thead{$-0.135^{***}$ \\ \textit{-2.978} \\ (0.2)} &         \thead{$0.032$ \\ \textit{1.12} \\ (86.2)} &  \thead{$-0.028$ \\ \textit{-0.915} \\ (21.1)} &        \thead{$0.231$ \\ \textit{1.197} \\ (88.9)} \\
				& \textbf{1.0} &     \thead{$-0.003$ \\ \textit{-0.184} \\ (41.1)} &      \thead{$-0.003$ \\ \textit{-0.133} \\ (45.0)} &  \thead{$-0.067^{***}$ \\ \textit{-5.079} \\ (0.0)} &   \thead{$-0.057^{***}$ \\ \textit{-2.93} \\ (0.4)} &    \thead{$-0.596^{*}$ \\ \textit{-1.644} \\ (7.3)} &        \thead{$0.017$ \\ \textit{0.106} \\ (31.3)} &  \thead{$-0.035$ \\ \textit{-1.065} \\ (15.2)} &    \thead{$1.256^{*}$ \\ \textit{2.082} \\ (94.9)} \\
				& \textbf{1.5} &  \thead{$3.352^{**}$ \\ \textit{3.909} \\ (95.3)} &      \thead{$-0.002$ \\ \textit{-0.003} \\ (78.8)} &       \thead{$-0.636$ \\ \textit{-0.817} \\ (20.6)} &             \thead{$0.0$ \\ \textit{0.0} \\ (86.9)} &       \thead{$-0.063$ \\ \textit{-0.246} \\ (33.1)} &   \thead{$0.462^{**}$ \\ \textit{1.276} \\ (98.5)} &  \thead{$-0.099$ \\ \textit{-0.107} \\ (79.4)} &        \thead{$0.121$ \\ \textit{0.273} \\ (83.3)} \\
				& \textbf{2.5} &       \thead{$0.018$ \\ \textit{0.138} \\ (70.4)} &      \thead{$-0.001$ \\ \textit{-0.012} \\ (58.7)} &       \thead{$-0.008$ \\ \textit{-0.186} \\ (45.3)} &        \thead{$-0.01$ \\ \textit{-0.224} \\ (43.8)} &       \thead{$-0.007$ \\ \textit{-0.099} \\ (49.6)} &        \thead{$0.008$ \\ \textit{0.078} \\ (63.0)} &  \thead{$-0.007$ \\ \textit{-0.081} \\ (51.0)} &        \thead{$0.027$ \\ \textit{0.177} \\ (66.5)} \\
				& \textbf{5.0} &     \thead{$-0.045$ \\ \textit{-0.039} \\ (67.8)} &      \thead{$-0.001$ \\ \textit{-0.016} \\ (57.7)} &         \thead{$0.001$ \\ \textit{0.034} \\ (52.8)} &       \thead{$-0.003$ \\ \textit{-0.068} \\ (49.9)} &       \thead{$-0.001$ \\ \textit{-0.013} \\ (51.8)} &         \thead{$0.003$ \\ \textit{0.03} \\ (61.7)} &  \thead{$-0.004$ \\ \textit{-0.048} \\ (52.0)} &        \thead{$0.019$ \\ \textit{0.154} \\ (62.8)} \\
				\cline{1-10}
				\multirow{5}[30]{*}{\textbf{0.15}} & \textbf{0.5} &     \thead{$-0.002$ \\ \textit{-0.187} \\ (39.2)} &      \thead{$-0.013$ \\ \textit{-1.066} \\ (13.7)} &    \thead{$-0.02^{**}$ \\ \textit{-1.995} \\ (2.9)} &    \thead{$-0.018^{*}$ \\ \textit{-1.462} \\ (7.8)} &       \thead{$-0.025$ \\ \textit{-0.831} \\ (20.5)} &        \thead{$0.027$ \\ \textit{1.179} \\ (87.5)} &  \thead{$-0.008$ \\ \textit{-0.321} \\ (38.7)} &   \thead{$0.114^{**}$ \\ \textit{2.131} \\ (97.8)} \\
				& \textbf{1.0} &     \thead{$-0.006$ \\ \textit{-0.454} \\ (31.0)} &      \thead{$-0.014$ \\ \textit{-0.947} \\ (16.0)} &   \thead{$-0.04^{***}$ \\ \textit{-3.602} \\ (0.0)} &   \thead{$-0.032^{**}$ \\ \textit{-2.332} \\ (2.2)} &  \thead{$-0.094^{***}$ \\ \textit{-2.525} \\ (0.1)} &        \thead{$0.031$ \\ \textit{1.199} \\ (88.5)} &  \thead{$-0.014$ \\ \textit{-0.476} \\ (33.2)} &     \thead{$0.109^{*}$ \\ \textit{1.48} \\ (91.9)} \\
				& \textbf{1.5} &     \thead{$-0.008$ \\ \textit{-0.511} \\ (27.9)} &       \thead{$-0.01$ \\ \textit{-0.587} \\ (27.2)} &  \thead{$-0.052^{***}$ \\ \textit{-4.379} \\ (0.0)} &  \thead{$-0.043^{***}$ \\ \textit{-2.729} \\ (0.8)} &  \thead{$-0.146^{***}$ \\ \textit{-2.925} \\ (0.5)} &        \thead{$0.035$ \\ \textit{1.202} \\ (88.6)} &  \thead{$-0.026$ \\ \textit{-0.838} \\ (23.5)} &        \thead{$0.287$ \\ \textit{1.075} \\ (83.2)} \\
				& \textbf{2.5} &  \thead{$3.312^{**}$ \\ \textit{5.786} \\ (97.9)} &      \thead{$-0.046$ \\ \textit{-0.106} \\ (58.8)} &        \thead{$-0.115$ \\ \textit{-0.25} \\ (26.4)} &     \thead{$0.19^{**}$ \\ \textit{0.729} \\ (96.8)} &        \thead{$-0.04$ \\ \textit{-0.224} \\ (40.0)} &   \thead{$0.354^{**}$ \\ \textit{2.596} \\ (98.5)} &  \thead{$-0.026$ \\ \textit{-0.037} \\ (83.4)} &         \thead{$0.016$ \\ \textit{0.04} \\ (71.7)} \\
				& \textbf{5.0} &       \thead{$0.005$ \\ \textit{0.021} \\ (68.0)} &      \thead{$-0.001$ \\ \textit{-0.018} \\ (57.8)} &         \thead{$0.001$ \\ \textit{0.022} \\ (52.4)} &       \thead{$-0.003$ \\ \textit{-0.076} \\ (49.4)} &        \thead{$-0.001$ \\ \textit{-0.02} \\ (51.6)} &        \thead{$0.003$ \\ \textit{0.034} \\ (61.6)} &  \thead{$-0.004$ \\ \textit{-0.051} \\ (52.1)} &        \thead{$0.019$ \\ \textit{0.155} \\ (63.2)} \\
				\cline{1-10}
				\multirow{5}[30]{*}{\textbf{0.25}} & \textbf{0.5} &       \thead{$0.002$ \\ \textit{0.178} \\ (53.0)} &      \thead{$-0.011$ \\ \textit{-1.095} \\ (12.6)} &        \thead{$-0.008$ \\ \textit{-0.94} \\ (18.1)} &         \thead{$0.005$ \\ \textit{0.535} \\ (68.8)} &         \thead{$0.031$ \\ \textit{1.314} \\ (88.9)} &    \thead{$0.026^{*}$ \\ \textit{1.283} \\ (90.2)} &      \thead{$0.0$ \\ \textit{0.016} \\ (50.9)} &    \thead{$0.11^{***}$ \\ \textit{2.54} \\ (99.5)} \\
				& \textbf{1.0} &      \thead{$-0.001$ \\ \textit{-0.07} \\ (45.0)} &   \thead{$-0.016^{*}$ \\ \textit{-1.408} \\ (7.9)} &       \thead{$-0.012$ \\ \textit{-1.333} \\ (10.1)} &       \thead{$-0.007$ \\ \textit{-0.633} \\ (25.9)} &          \thead{$0.02$ \\ \textit{0.763} \\ (78.5)} &        \thead{$0.028$ \\ \textit{1.241} \\ (88.6)} &   \thead{$-0.006$ \\ \textit{-0.23} \\ (41.8)} &  \thead{$0.133^{***}$ \\ \textit{2.703} \\ (99.6)} \\
				& \textbf{1.5} &     \thead{$-0.005$ \\ \textit{-0.393} \\ (32.8)} &   \thead{$-0.018^{*}$ \\ \textit{-1.488} \\ (7.3)} &   \thead{$-0.03^{***}$ \\ \textit{-3.118} \\ (0.2)} &    \thead{$-0.017^{*}$ \\ \textit{-1.431} \\ (8.5)} &  \thead{$-0.075^{***}$ \\ \textit{-2.418} \\ (0.5)} &        \thead{$0.028$ \\ \textit{1.168} \\ (87.5)} &   \thead{$-0.01$ \\ \textit{-0.386} \\ (35.9)} &   \thead{$0.117^{**}$ \\ \textit{2.012} \\ (97.2)} \\
				& \textbf{2.5} &     \thead{$-0.012$ \\ \textit{-0.844} \\ (19.3)} &      \thead{$-0.018$ \\ \textit{-1.101} \\ (13.3)} &  \thead{$-0.047^{***}$ \\ \textit{-4.508} \\ (0.0)} &   \thead{$-0.04^{***}$ \\ \textit{-2.767} \\ (0.9)} &  \thead{$-0.162^{***}$ \\ \textit{-3.334} \\ (0.5)} &        \thead{$0.033$ \\ \textit{1.148} \\ (87.8)} &  \thead{$-0.012$ \\ \textit{-0.387} \\ (35.7)} &        \thead{$0.426$ \\ \textit{1.187} \\ (81.7)} \\
				& \textbf{5.0} &       \thead{$0.014$ \\ \textit{0.131} \\ (68.0)} &      \thead{$-0.002$ \\ \textit{-0.029} \\ (58.1)} &       \thead{$-0.002$ \\ \textit{-0.041} \\ (50.6)} &       \thead{$-0.005$ \\ \textit{-0.123} \\ (47.3)} &       \thead{$-0.004$ \\ \textit{-0.058} \\ (50.5)} &        \thead{$0.005$ \\ \textit{0.049} \\ (61.2)} &  \thead{$-0.006$ \\ \textit{-0.072} \\ (51.3)} &        \thead{$0.019$ \\ \textit{0.126} \\ (64.1)} \\
				\cline{1-10}
				\multirow{5}[30]{*}{\textbf{0.35}} & \textbf{0.5} &       \thead{$0.006$ \\ \textit{0.617} \\ (68.4)} &      \thead{$-0.008$ \\ \textit{-0.777} \\ (22.0)} &        \thead{$-0.007$ \\ \textit{-0.94} \\ (18.4)} &      \thead{$0.01^{*}$ \\ \textit{1.275} \\ (90.2)} &     \thead{$0.036^{*}$ \\ \textit{1.707} \\ (92.6)} &   \thead{$0.029^{**}$ \\ \textit{1.656} \\ (96.0)} &    \thead{$0.005$ \\ \textit{0.265} \\ (59.1)} &        \thead{$0.084$ \\ \textit{2.293} \\ (99.0)} \\
				& \textbf{1.0} &       \thead{$0.004$ \\ \textit{0.418} \\ (61.1)} &      \thead{$-0.013$ \\ \textit{-1.266} \\ (10.3)} &        \thead{$-0.01$ \\ \textit{-1.187} \\ (12.7)} &     \thead{$0.012^{*}$ \\ \textit{1.393} \\ (92.5)} &     \thead{$0.034^{*}$ \\ \textit{1.524} \\ (90.8)} &    \thead{$0.026^{*}$ \\ \textit{1.398} \\ (93.2)} &  \thead{$-0.003$ \\ \textit{-0.131} \\ (45.1)} &  \thead{$0.094^{***}$ \\ \textit{2.357} \\ (99.3)} \\
				& \textbf{1.5} &       \thead{$0.003$ \\ \textit{0.252} \\ (54.9)} &   \thead{$-0.017^{*}$ \\ \textit{-1.575} \\ (6.0)} &   \thead{$-0.011^{*}$ \\ \textit{-1.337} \\ (10.0)} &          \thead{$0.005$ \\ \textit{0.59} \\ (72.4)} &          \thead{$0.033$ \\ \textit{1.42} \\ (89.4)} &        \thead{$0.024$ \\ \textit{1.214} \\ (89.9)} &  \thead{$-0.007$ \\ \textit{-0.301} \\ (38.2)} &  \thead{$0.107^{***}$ \\ \textit{2.419} \\ (99.4)} \\
				& \textbf{2.5} &     \thead{$-0.002$ \\ \textit{-0.211} \\ (39.1)} &  \thead{$-0.023^{**}$ \\ \textit{-1.948} \\ (3.7)} &  \thead{$-0.028^{***}$ \\ \textit{-3.198} \\ (0.2)} &        \thead{$-0.007$ \\ \textit{-0.72} \\ (22.5)} &  \thead{$-0.083^{***}$ \\ \textit{-3.044} \\ (0.1)} &        \thead{$0.023$ \\ \textit{1.011} \\ (83.6)} &  \thead{$-0.009$ \\ \textit{-0.377} \\ (36.2)} &  \thead{$0.145^{***}$ \\ \textit{2.504} \\ (99.5)} \\
				& \textbf{5.0} &     \thead{$-0.011$ \\ \textit{-0.027} \\ (69.1)} &        \thead{$-0.042$ \\ \textit{-0.1} \\ (55.9)} &       \thead{$-0.027$ \\ \textit{-0.146} \\ (38.2)} &        \thead{$-0.02$ \\ \textit{-0.341} \\ (39.5)} &       \thead{$-0.026$ \\ \textit{-0.127} \\ (44.2)} &  \thead{$0.479^{***}$ \\ \textit{1.908} \\ (99.3)} &   \thead{$-0.072$ \\ \textit{-0.13} \\ (49.3)} &      \thead{$-0.002$ \\ \textit{-0.006} \\ (66.7)} \\
				\bottomrule
		\end{tabular}}
	\caption{\scriptsize \textbf{Nonlinear DebtRank}.
	This table reports data about the relative equity loss of the system $H$ with the nonlinear DebtRank valuation function (Figure \ref{fig:nonlin_aggr}). Each row corresponds to a combination of the initial shock $\lambda$ and nonlinearity parameter $\alpha$ while the columns correspond to the yearly networks. 
    The top element from each cell corresponds to the relative deviation of $H$ for e-MID data from its SDECM average; the middle elements in \textit{italics} correspond to the difference between e-MID and SDECM average divided by SDECM standard deviation; the bottom element in parenthesis reports the percentile position of e-MID with respect to the distribution of $H$ generated by SDECM. 
    *, ** and *** denote the significance of the deviation and highlight if e-MID's $H$ is either in the top or bottom 10\%, 5\% or 1\% of the distribution generated by SDECM.
	}
	\label{table:nonlinear}
	\end{table}

\end{document}